\documentclass[10pt,preprint]{aastex}

\shorttitle{Star formation in the LMC}
\shortauthors{Kawamura et al. }

\begin{document}
\title{The Second Survey of the Molecular Clouds in the Large Magellanic Cloud by NANTEN. II. Star Formation}
\author{Akiko Kawamura\altaffilmark{1},
Yoji Mizuno\altaffilmark{1},
Tetsuhiro Minamidani\altaffilmark{1,2},\\
Miroslav D.\ Fillipovi\'{c},\altaffilmark{3},
Lister Staveley-Smith\altaffilmark{4},
Sungeun Kim\altaffilmark{5},\\
Norikazu Mizuno\altaffilmark{1},
Toshikazu Onishi\altaffilmark{1,6},
Akira Mizuno\altaffilmark{7}, \&
Yasuo Fukui\altaffilmark{1}
}

\altaffiltext{1}{Department of Astrophysics, Nagoya University, 
Furocho, Chikusaku, Nagoya 464-8602, Japan}
\altaffiltext{2}{Department of Physics, Faculty of Science, 
Hokkaido University, N10W8, Kita-ku, Sapporo 060-0810, Japan}
\altaffiltext{3}{University of Western Sidney, Penrith South DC, NSW 1797, Australia}
\altaffiltext{4}{School of Physics, M013, University of Western Australia, 
35 Stirling Highway, Crawley, WA 6009, Australia}
\altaffiltext{5}{Astronomy \& Space Science Department, 
Sejong University, 98 Kwangjin-gu, Kunja-dong, Seoul, 143-747, Korea}
\altaffiltext{6}{Department of Physical Science, Osaka Prefecture University, Gakuen 1-1, Sakai, Osaka 599-8531, Japan}
\altaffiltext{7}{Solar-terrestrial Environment Laboratory, Nagoya University, 
Furocho, Chikusaku, Nagoya 464-8601, Japan}
\email{kawamura@a.phys.nagoya-u.ac.jp}

\begin{abstract}
We studied star formation activities in the molecular clouds in the Large Magellanic Cloud. We have utilized the second catalog of 272 molecular clouds obtained by NANTEN to compare the cloud distribution with signatures of massive star formation including stellar clusters, and optical and radio HII regions. We find that the molecular clouds are classified into three types according to the activities of massive star formation; Type I shows no signature of massive star formation, Type II is associated with relatively small HII region(s) and Type III with both HII region(s) and young stellar cluster(s). The radio continuum sources were used to confirm that Type I GMCs do not host optically hidden HII regions. These signatures of massive star formation show a good spatial correlation with the molecular clouds in a sense they are located within $\sim$100 pc of the molecular clouds. Among possible ideas to explain the GMC Types, we favor that the Types indicate an evolutionary sequence; i.e., the youngest phase is Type I, followed by Type II and the last phase is Type III, where the most active star formation takes place leading to cloud dispersal. The number of the three types of GMCs should be proportional to the time scale of each evolutionary stage if a steady state of massive star and cluster formation is a good approximation. By adopting the time scale of the youngest stellar clusters, 10 Myrs, we roughly estimate the timescales of Types I, II and III to be 6 Myrs, 13 Myrs and 7 Myrs, respectively, corresponding to a lifetime of 20--30 Myrs for the GMCs with a mass above the completeness limit, $5 \times 10^{4} M_\sun$.
\end{abstract}


\section{Introduction}
The Magellanic Clouds can be observed in more detail than any other extra galaxies at any wavelengths because of the proximity. The relatively face-on location of the LMC enables us to obtain a complete sample of astronomical objects with less contamination compared with the Galaxy. Studies of the LMC provide invaluable information on our understanding of the galaxies in various aspects, including the properties of the ISM, evolution of molecular clouds and star formation. 

The environments, such as  metallicity, in the LMC are different from those in the Galaxy (e.g., $Z \sim 1/2 Z_\odot$, Dufour 1984 and taking into account the revision of the Solar abundance by Asplund et al.\ 2004).
Star formation activities are also different. Stellar clusters called ``populous clusters", which are self-gravitating like Galactic globular clusters, are found by photometric studies (e.g., Hodge 1961; van den Bergh 1981). Their masses are $\sim$ 10$^4$--10$^5 \, M_{\Sol}$, which are smaller than those of the Galactic globular clusters but larger than those of the Galactic open clusters by an order of magnitude (Kumai, Basu, \& Fujimoto 1993; Hunter  et al. 2003). It is notable that more than a hundred of the populous clusters are significantly younger, i.e., a few to 100 Myr, than the Galactic globular clusters and some are still forming at present, such as R136 in 30 Dor nebula (e.g., Massey \& Hunter 1998). 
This suggests that the formation process of globular-like rich clusters can be studied through the observations of the young clusters and ISM properties in the LMC. 
To date, optical indicators of the massive star formation or cluster formation, such as H~{\sc ii} regions and stellar clusters, have been studied in the large area of the LMC (e.g., Henize 1956; Davies et al.\ 1976, hereafter DEM; Kennicutt \& Hodge 1986, hereafter KH; Bica et al.\ 1996). 
DEM identified 357 H$\alpha$ emission nebulae, and KH measured the H$\alpha$ flux of 240 H~{\sc ii} regions. Regarding the stellar clusters, Bica et al.\ (1996) cataloged 624 clusters, and classified them into 8 types according to their colors. 

The first complete map of the molecular gas in the LMC was obtained by Cohen et al.\ (1988) in $^{12}$CO (1-0) with the southern CfA 1.2 m telescope at CTIO. However, the survey was limited by the low spatial resolution, $8.\arcmin8 $ corresponding to 130 pc at the distance of the LMC.
High-resolution CO observations of selected regions, especially toward well-known active star forming regions, for example, 30 Dor, N11, N159, by the SEST 15 m telescope have been performed in the LMC (e.g., Israel et al.\ 1986; Johansson et al.\ 1994; Caldwell \& Kutner 1996; Kutner et al.\ 1997; Johansson et al.\ 1998; Israel et al.\ 2003). These observations revealed detailed structure and properties of the molecular gas of the individual star forming regions at a linear resolution of less than 10 pc, although they are limited in spatial coverage, about one square degree.  

Recently, Fukui et al.\ (2008, hereafter, ``Paper I'') made a second survey of the molecular gas in the LMC by a 4m-telescope, NANTEN, at Las Campanas Observatory, Chile. This survey was carried out in $^{12}$CO (1-0) with resolution of 2 arcmin grid spacing with the half-power beam width of $2\arcmin.6$ and covered $\sim 30$ deg$^2$ (Fukui et al. 2008). The resolution of the survey was high enough to resolve giant molecular clouds and enabled us to cover a large region efficiently. The molecular clouds with a completeness limit of $5 \times 10^4 M_{\sun}$ in mass are identified in nearly the entire region where the current massive star and cluster formation is on-going, and 272 molecular clouds are cataloged.

In this paper, we present the results from comparisons of the GMCs identified by NANTEN (Paper I) with classical H II regions and optically identified stellar clusters and discuss cloud evolution. Recent surveys of the Magellanic Clouds by the IR satellites, like Spitzer (e.g., Meixner et al. 2006, `Surveying the Agents of a Galaxy's Evolution') and AKARI (e.g., Ita et al. 2008; Murakami et al. 2008) have been strong tools to identify younger, and lower mass YSOs (Whitney et al. 2008). The comparisons of these YSOs and the GMCs are found in elsewhere (Indebetouw et al. 2008; 
Onishi et al. 2009, in preparation).

\section{Molecular clouds, HII regions, and young clusters}

\subsection{Molecular clouds identified by the 2nd NANTEN survey}
A survey of the molecular clouds was carried out in $^{12}$CO ($J$=1--0) by NANTEN, a 4 m radio telescope of Nagoya University at Las Campanas Observatory, Chile (Paper I). The observed region is about 30 square degrees and covers the region where the CO emission was detected by the NANTEN first survey (e.g., Fukui et al. 1999; Mizuno et al.\ 2001). The observed grid spacing was 2$\arcmin$, corresponding to $\sim$ 30 pc at a distance of the LMC, 50 kpc, with a 2.6$\arcmin$ half-power beam width at 115 GHz. The spectral intensities were calibrated by employing the standard room-temperature chopper wheel technique (Kutner \& Ulich 1981). An absolute intensity calibration was made by observing Orion-KL [R.A. (B1950) $= 5^{h}32^{m}47^{s}$. 0, Decl. (B1950) $= -5\arcdeg 24\arcmin 21\arcsec$] by assuming its absolute temperature, $T_{\rm R}^{*}$, to be 65 K. The rms noise fluctuations were about 0.07 K at a velocity resolution of 0.65 km s$^{-1}$ with about 3 minutes integration for an on-position. The typical $3 \sigma$ noise level of the velocity-integrated intensity was about 1.2 K km s$^{-1}$
(Figure 1). 

Fukui et al.\ (2008) identified 272 molecular clouds of which 230 are detected at more than two observing positions (hereafter, "GMCs", in this paper) by using cloud identifying algorithm, cprops (Rosolowsky \& Leroy 2006). The radius and virial mass of the clouds range from 10 to 220 pc, and $9 \times 10^{3}$ to $9 \times 10^{6} M_{\odot}$, respectively.
The CO luminosity and virial mass of the GMCs show a good correlation,
and a conversion factor, $X_{\rm CO}$, from a CO intensity to
an H$_2$ column density was derived to be (7 $\pm 2$) $\times 10^{20}$ cm$^{-2}$ (K km s$^{-1}$)$^{-1}$ by assuming virial equilibrum. 
The sensitivity in $N$(H$_2$), then, corresponds to $N$(H$_2$) = $8 \times 10^{20}$ cm$^{-2}$ and the range of mass of the GMCs is
$2 \times 10^4 M_{\sun}$ to $7 \times 10^6 M_{\sun}$, respectively. The details of the observations and the method to identify the GMCs are found in Paper I and Rosolowsky \& Leroy (2006).

\subsection{Young astronomical objects associated with the molecular clouds}
In order to identify the current massive star and cluster forming molecular clouds, H~{\sc ii} regions and young stellar clusters were searched for in the published catalogs. The association of these objects and the molecular clouds were determined. 

Henize (1956) and Davies et al.\ (1976) cataloged more than 300 H$\alpha$ emission nebulae in the LMC. The estimated diameters of the HII regions range from $\sim$ 10 pc to $\sim$ 400 pc. In addition, an extensive H$\alpha$ photometry of the 240 HII regions was carried out by Kennicutt \& Hodge (1986). Figure 2 shows luminosity distribution of these HII regions, indicating that the current sample of HII regions includes those with H$\alpha$ luminosities as faint as $\sim 10^{36}$ erg s$^{-1}$ at the faint end. This shows that the sensitivity of the survey is high enough to detect the Orion nebula, $\sim 4 \times 10^{36}$ erg s$^{-1}$, (Gebel 1968) at the distance of the LMC. The Orion Nebula is ionized primarily by a single O7 star together with at least B0.5, B3 and B0.5 (O9.5) stars (Goudis 1982). This faint end of the luminosities is comparable to those of small Galactic HII regions that can be ionized by a single O9 star (e.g., Pagnia \& Ranieri 1973). Detailed studies of several faint HII regions in the LMC show that most of the HII regions are ionized primarily by a star as massive as B0 accompanied by several other massive but cooler stars, except for two ionized by a single mid-O star and B0 star (Wilcots 1994). It is also suggested that the faint HII regions in the catalogs are ionized by a single O star and those luminous ones ($L \gtrsim 5 \times 10^{37}$ erg s$^{-1}$) are by OB associations (Kennicutt \& Hodge 1986).

A mosaic image of the 1.4 GHz continuum emission observed with the Parkes Telescope (Filipovic et al.\ 1995) and the ATCA are combined to present the thermal and nonthermal radio emission of the LMC covering $10.8\arcdeg \times 12.3\arcdeg$ (e.g., Filipovic et al. 1998b; Hughes et al.\ 2007). This image was also used to determine if there are any HII regions escaped from optical identification due to the extinction by molecular clouds. The angular resolution of the data is $40 \arcsec$, corresponding to $\sim 10$ pc at a distance of the LMC. Thus one should keep in mind that a size of the HII regions of this study is larger than $\sim 10$ pc.

The LMC has more than 6500 stellar clusters and OB associations (Bica et al.\ 1999). Photometric properties of about 10 \% of the brightest ones were measured to derive their masses and ages (Bica et al. 1996). In this paper, we use these clusters with known properties to study the formation of cluster and OB association in the molecular clouds. These 120 OB associations and 504 stellar clusters are classified into 10 types of different ages (SWB0--VII) (Bica et al.\ 1996) from their $U-B$ and $B-V$ indices. In table 1, the numbers of these objects according to their ages in the area covered with the 2nd NANTEN survey are shown. We shall call these objects ``clusters'' throughout this paper and we should note that this includes both clusters and OB associations.

\section{Molecular clouds and association with HII regions and young clusters}
\subsection{Overall distribution of the molecular clouds, HII regions and clusters}

Figures 1, 3a, and 3b show the distribution of the H$\alpha$ emission (Kim et al.\ 1999), young clusters (younger than 10 Myr), and clusters older than 10 Myr together with the molecular clouds, respectively. 

Figure 1 shows that the overall distribution of the H$\alpha$ local peaks and GMCs are well coincident to each other, while the extent of the individual H$\alpha$ emitting regions and the size of the GMCs are different. 
Almost all the luminous HII regions with high flux densities, like 30 Dor to N 159, N 11, N 63/N 64 complexes, N 44 and N 206 are associated with GMCs, while fainter and diffuse extended H$\alpha$ emission show less degrees of association. Some luminous HII regions, like N 51, are associated with only small molecular clouds, which may indicate molecular gas dissipation after formation of massive HII regions. It is also noted that there are several GMCs not associated with H$\alpha$ emission (see also section 4.1)

Figure 3a shows that young clusters are often found at, or near the peak of the GMCs. Some of the young clusters associated with GMCs are forming groups with a few to ten clusters. These active cluster forming regions, like N 159, N 11, and N 44, are found especially toward bright HII regions.  
Note that a number of clusters in the northeastern region in the LMC and some in the west of 30 Dor are isolated from molecular clouds. These groups of clusters are located inside large cavities of H$ \alpha$ emission (see also Fig. 1), two of the supergiant shells, LMC 4 and LMC 3, respectively (Meaburn 1980).

Compared with the youngest group of clusters, the older clusters have less degree of association with the GMCs (Figure 3b). The second youngest group of clusters, SWB I, is presented in blue crosses in Fig. 3b, showing that the correlation with the molecular clouds is low. Nevertheless there are several regions where the number of SWB I clusters is enhanced, for example, inside a supergiant shell, LMC 4, and the south of 30 Dor. It is interesting to note that the SWB 0 clusters are also gathered in these regions, suggesting that the distribution of SWB 0 and I clusters still retain the information of their formation sites. On the other hand, the older clusters, SWB II -- VII, are distributed more uniformly over the galaxy showing no trend of association with the GMCs. It is to be noted here that the older clusters are distributed more widely over the galaxy compared with the region where the current star formation is observed (Bica et al.\ 1996)

The overall distribution of HII regions, young clusters and older clusters indicates that the GMCs are the sites of current massive star formation and of cluster formation.
To show the associated object more clearly, we present close-up views of the GMCs and the young objects in Figures 4 and 5a--f (Figure 4 as a guide). These panels confirm that the HII regions and young clusters are often associated with molecular clouds well, while the older clusters are distributed randomly.  

In the following, we report some individual active star forming regions in detail. In the N 11 (DEM 41) region, the young clusters are found in a bright, large complex of HII regions, while they are not exactly found toward the GMCs (Figure 5a).  Only small clouds are found at the outer edge of N 11 nebulae with young clusters near the edge of the molecular clouds. The cluster at the center of N 11, LH 9, is neither associated with H$\alpha$ emission peaks nor molecular clouds. This suggests that the parent cloud of LH 9 has been dissipated, and that triggered star formation occurred at the outer region as discussed by for example, Israel et al. (2003a). 
A group of HII regions are found near the western edge of the bar in the N 79/N 83 region (Figure 5b). In contrast with the N 11 region, young clusters associated with the GMCs and bright H$\alpha$ emission are found near the peak of the GMCs in this region, and the young clusters without bright HII region are at the center of diffuse shell-like H$\alpha$ emission. In Figure 5b, there are a few GMCs hosting no H II regions, one of which is at the western end of the N79/N 83 region.

Small groups of young clusters are found near the peak of a massive GMC in N 44 (DEM158, 160, 166, 167, 169) as well as small GMCs along the supergiant shell LMC 4 (Figure 5c). On the other hand, only one or two clusters are associated with a GMC near the LMC bar (Figure 5d). One may speculate that clusters are formed in groups in the massive GMCs of $M \sim 10^6 M_{\sun}$ or near a supergiant shell.  It is also noted that bright HII region, the N 51 complex, is associated with a very small molecular cloud, maybe a remnant of the parent cloud of this bright HII region.

Figure 5f shows the 30 Dor (DEM 263) region to the molecular ridge including active star forming site, N 159 (DEM 271, 272), and the Arc region. The most remarkable feature in H$\alpha$ emission is a bright complex of HII regions in 30 Dor, while massive molecular clouds are found not exactly toward 30 Dor but extending to the south as already noted by several authors (e.g. Cohen et al. 1988; Indebetouw et al. 2008; Paper I).  Most of the youngest clusters of $\tau <$ 10 Myr are associated with the GMCs or found near the GMCs. On the other hand, the clusters of 10 $< \tau <$ 30 Myr are away from the GMCs and mostly found between 30 Dor and N 159 where the bright H$\alpha$ emission from 30 Dor is extended but only small molecular clouds are detected.  
The most active current star formation site in this figure is a well-known region, N 159, where young clusters as well as a number of HII regions are found at or near the peaks of the molecular ridge. The detailed studies of star formation activities and molecular clouds are carried out by several authors (e.g., Johansson et al. 1998; Minamidani et al. 2008; Indebetouw et al. 2008; Mizuno et al. 2009). In the south of N159, only smaller HII regions are associated with the molecular ridge. These results indicate that the southern region may be younger than the North. 

There are several other regions with active star formation 
in Figure 5f, N 148, N 180, N 206 and N 214, where groups of HII regions, some of which contain a young cluster, are associated with GMCs. It is interesting to note that these groups of HII regions are mostly found off from the peak of the GMCs. It should be noted that the NANTEN beam is not capable of resolving the individual local peaks of the GMC and the parent core of the group of HII regions may have escaped from detection. Nevertheless, the positional offset of the peak of the GMC and the group of HII regions indicate the dissipation of the molecular gas by active star formation. It is also to be noted that there are several GMCs, especially at southern edge of this figure without significant H$\alpha$ emission or clusters. 

Figure 6 shows the distribution of the projected separations of H II regions and stellar clusters from the nearest CO emission with an integrated intensity above 1.2 K km s$^{-1}$ (3$\sigma$ noise level). The lines in Fig. 6 represent the frequency distribution expected if the same number of the H~{\sc ii} regions or clusters are distributed at random in the oserved area. It is clearly shown by eye that the distribution of the youngest clusters with an age smaller than 10 Myrs, i.e., SWB 0 (Bica et al.\ 1996), and the H II regions are sharply peaked within 100 pc of CO emission exhibiting strong spatial correlations. On the other hand, the older clusters, SWB I  ($\tau > 10$ Myr) or older, show much weaker or no correlation. This Figure again indicates that the HII regions and SWB 0 clusters are well associated with molecular clouds as well as that rapid cloud dissipation after cluster formation. 

Hereafter, we will focus on the HII regions and SWB 0 clusters in the discussion and call SWB 0 cluster as ``young cluster" unless otherwise stated. 

\subsection{Determination of association of individual molecular clouds with HII regions and young clusters}
\subsubsection{Determination of association of the HII regions and young clusters}
We have determined the association of individual HII regions and young clusters with the molecular clouds by the following criteria: 1) for clusters, if the extent of a cluster is overlapped with the boundary of a molecular cloud, 2) for the H II regions, if the H$\alpha$-emitting region is overlapped with the boundary of a molecular cloud.

In total, 97 out of 137 young clusters are found to be associated with the molecular clouds. In addition, those clusters are all associated with HII regions. These confirms that the clusters associated with the molecular clouds are young and contain massive stars. For reference, we have also determined the association of the SWB I clusters and the molecular clouds. As a result, only 15 out of 122 clusters are found to be associated with the molecular clouds in the survey area of NANTEN. This result suggests that the molecular clouds start to be dissipated while the clusters are still in the SWB 0 phase. This scenario is consistent with the result from Fukui et al.\ (1999) and Yamaguchi et al. (2001b). 

\subsubsection{Determination of association of radio continuum sources toward the molecular clouds without optically identified HII regions}

As presented in the previous sections, there are a number of molecular clouds without HII regions or young clusters. 
Clusters and H~{\sc ii} regions may be hidden behind the molecular clouds by chance. For the H II regions, we have also compared the distribution of the molecular clouds and the point sources from the ATCA+Parkes combined continuum emission (Filipovic et al.\ 1995; Dickel et al.\ 2005; Hughes et al.\ 2007) to search for such hidden H~{\sc ii} regions. If an H~{\sc ii} region is behind a GMC, the H~{\sc ii} region cannot be seen optically due to extinction, but should be observed as a radio source. 

A mosaic image of the 1.4 GHz continuum emission observed with the Parkes Telescope (Filipovic et al.\ 1995) and the ATCA are combined to present the thermal and nonthermal radio emission of the LMC covering $10.8\arcdeg \times 12.3\arcdeg$ (e.g., Filipovic et al. 1998b; Hughes et al.\ 2007). A catalog of sources identified from this image will be presented in elsewhere (Filipovic et al. 2008, in preparation). In this work, first, 1.4 GHz continuum emission was examined carefully toward the molecular clouds without optical HII region. Then if we see any point sources at 1.4 GHz, we searched for corresponding 4.8 and 8.6 GHz sources (Dickel et al. 2005).  We found about seventy 1.4 GHz both point-like and extended sources toward the 72 molecular clouds; ring-like artifacts seen in the 1.4 GHz image near 30 Dor made it difficult to determine the emission toward four molecular clouds. Among these 1.4 GHz sources, only one 1.4 GHz source, ATCA J054308-710409, was also seen at 4.8 GHz and 8.6 GHz, indicating that most of the 1.4 GHz sources in these molecular clouds are background sources or SNRs. 

The 1.4 GHz source, ATCA J054308-710409, is found just at the edge of the GMC 224 and coincides with a radio source, LMC B0543-7105, identified at 4.75, 4.85 and 8.85 GHz by Fililpovic et al. (1995). Because it lies at the edge of the GMC, the absence of H$\alpha$ emission toward the source is not perhaps due to the extinction by the molecular cloud. We estimated a spectral index of this source, $\alpha$, defined as $S_{\nu} \sim \nu^{\alpha}$, where $S_{\nu}$ is the integrated flux density at frequency, $\nu$, to be $\alpha \sim -0.4$. Filipovic et al. (1998a) studied the spectral index of radio sources detected with the Parkes telescope and found that the known HII regions have rather flat spectra with spectral index of $\alpha = -0.15 \pm 0.31$, and the SNRs and background sources have steeper spectra, $\alpha = -0.43 \pm 0.19$ and $\alpha = -0.59 \pm 0.48$, respectively. These results show that ATCA J054308-710409 is also unlikely to be a hidden HII region and is more likely to be an SNR.

This comparison of the ATCA 1.4 GHz and the molecular clouds indicate that hidden HII regions of this size are unlikely to exist. Since all the young clusters found in the molecular clouds are associated with the H II regions, these results suggest that the hidden clusters are also unlikely to exist.

\subsection{Molecular cloud Types}

It was shown in Fukui et al.\ (1999) that the GMCs can be classified into three groups based on a sample of 55 GMCs with a mass ranging from $2 \times 10^{5} M_{\odot}$ to $3 \times 10^{6} M_{\odot}$; 1) starless GMCs, 2) those with small H{\small \,II} regions whose H$\alpha$ luminosity is less than $10^{37}$ erg s$^{-1}$, and 3) those with stellar clusters and large H{\small \,II} regions of H$\alpha$ luminosity greater than $10^{37}$ erg s$^{-1}$.

The current comparison of the molecular clouds with the clusters and HII regions gives a consistent result. Here we classify the molecular clouds into three types according to the association with {\bf massive} star formation activities; 

\begin{enumerate}
\item starless molecular clouds in the sense that they are not associated with HII regions or young clusters (Type I),
\item  molecular clouds with HII regions (Type II), and 
\item molecular clouds with HII regions and young clusters (Type III). 
\end{enumerate}

It should be noted that ``starless'' here means without star forming activities with stars more massive than early O star capable of ionizing HII regions, and it does not exclude the possibility of associated young low mass stars. Comparisons of the GMCs with young, low or intermediate mass stars are now possible by using recent results by the IR satellites, like Spitzer (e.g., Meixner et al. 2006, `Surveying the Agents of a Galaxy's Evolution') and AKARI (e.g., Ita et al.\ 2008; Murakami et al. 2008). The comparisons of these YSOs (Whitney et al. 2008) and the GMCs are found in elsewhere
(Indebetouw et al. 2008, Onishi et al. 2009, in preparation).

Table 2 lists the associated HII regions and young clusters for each molecular cloud.
Out of 272 molecular clouds, 72, 142 and 58 are found to be Type I, II, and III, respectively (see also Table 3).
Figures 7 to 9 present the examples of the molecular clouds of each Type. Examples are chosen from the most massive clouds from each type. It is interesting to note that the most massive Type I molecular clouds have a similar size and mass as those of Type II, while the number of massive Type I is less. To study the physical properties of the molecular clouds, one has to keep in mind that the completeness limit of the NANTEN survey of $M_{\rm CO} = 5 \times 10^4 M_{\sun}$. Table 3 also summarizes a number of GMCs with $M_{\rm CO} > 5 \times 10^4 M_{\sun}$ for each Type. Out of 191 GMCs, 46, 96 and 49 GMCs are found to be Type I, II, III, showing that about a half of them are Type II, and a quarter is Type I and Type III, respectively. 

\subsection{Distribution of the molecular clouds}

Figures 10a-f show the radial distribution of CO emission;
the number and the surface density, $\Sigma$
of the molecular clouds with Types I, II and III, respectively.
The surface density, $\Sigma$, is derived by integrating 
the CO luminosity within annuli spaced by 4$\arcmin$ 
and then divided by an area of the annuli.
The center used is 
$\alpha$(J2000)$=5^{h}17.6^{m}$, 
$\delta$(J2000)$=-69\arcdeg2^{'}$ 
determined from the kinematics of the HI observations by Kim et al.\ (1998). To see the angular distribution of the CO emission, the distribution of the clouds and the surface density, $\Sigma$, for each molecular cloud Type are also presented in Figures 11a--f. Here, the surface density, $\Sigma$, is derived by integrating CO luminosity over a sector with a $20\arcdeg$ width and then divided by an observed area of the sector. The CO luminosity to mass conversion is carried out by assuming a conversion factor, $X_{\rm CO}$ of $7 \times 10^{20}$ cm$^{-2}$(K km s$^{-1}$)$^{-1}$ (Paper I) for both Figures 10 and 11.

Figure 10 shows that the radial profile of the surface density decreases moderately along the galacto-centric distance for the Type II as is also seen in the nearby spiral galaxies (e.g., Wong \& Blitz 2002), while those for the Types II and Type III are rather flat with respect to the radial distance.
It is interesting to note that the number distribution and surface density show different radial profiles for Type I; the number increases at large radial distances but the surface density is relatively constant. This indicates that the more massive Type I GMCs are found at the large radial distances. It is also notable that there is a sharp enhancement of the number of the clouds around 1.5 kpc for Type II and III. This enhancement is due to the molecular ridge, N11, and N44. 
This enhancement is also seen in the angular distribution, especially at about $120 \arcdeg$ due to the molecular ridge.

\subsection{Physical Properties of the GMCs}

Cluster forming clouds provide us with precious information on understanding physical processes of cluster formation.  We shall examine the physical properties of the GMC Types. In this section, only the GMCs with mass, $M_{\rm CO} > 5 \times 10^4 M_{\sun}$ are considered. Figure 12 shows the distribution of the line width, size, and mass of the GMCs, respectively.  The upper panels are the GMCs not associated with the HII regions and the clusters, i.e., those with no sign of massive star formation (Type I).  The middle ones are the GMCs associated only with the HII regions, i.e., those showing massive star formation but no cluster formation (Type II). The lower are the GMCs associated with the HII regions and clusters, i.e., those actively forming stars and clusters (Type III). Table 4 summarizes a mean and standard deviation of the line widths, sizes and masses of each GMC Type.

Figure 12 and Table 4 indicate that a significant difference is not seen in the line width and size among the three Types, while the mass distribution shows a slight difference. The frequency distribution of mass of Type I and II shows a peak at $M_{\rm CO} \sim 10^5 M_{\sun}$ and then decreases as the mass becomes large, while that of the Type III is rather flat and the average mass of the Type III is larger than that of Type I and II. The small difference in the physical properties among three types suggests that the integrated properties of the GMCs at a scale of 30 -- 100 pc are not so sensitive to the local star formation activities until the last stage of cloud dissipation.

\section{Discussion}
\subsection{GMC Type I: GMCs without massive star formation}

Almost all the GMCs in the Solar vicinity are forming massive stars actively as indicated by associated HII regions and/or OB associations in addition to a number of young low-mass stars  
(Dame et al. 1987; Blitz 1993 and references therein). There are only two GMCs which show no signs of massive star formation in the solar neighborhood among $\sim 20$ GMCs; the Maddalena's cloud (Maddalena \& Thaddeus 1985) and ON-1 cloud complex (Israel \& Wootten 1982). The reason for not forming massive stars may be either it is at a very young stage prior to massive star formation (Maddalena \& Thaddeus 1985; Israel \& Wootten 1986) or that it is in a late stage after active star formation (Lee, Snell, \& Dickman 1994).  

A large number of starless GMCs in the LMC suggests that the timescale in star formation is significantly longer in the LMC than in the Galaxy. The ionization degree in a molecular cloud is likely determined by the far-ultraviolet (FUV) photons of stellar radiation fields (McKee 1989; Nozawa et al.\ 1991). In the LMC, the FUV flux is several times higher (Israel et al. 1986) and the dust extinction is smaller by a factor of 3--4 for a given gaseous column density (Koornneef 1982). Since the time scale of the diffusion of magnetic field is proportional to the ionization degree (Spitzer 1978), the contraction of cloud may be slowed down by the magnetic field. In addition, the cooling rate via molecular and dust emission is expected to be smaller in the LMC than in the Galaxy, helping a star formation activity to slow down.
Higher ionization degree and smaller cooling rate are basically the consequences of lower metallicity in the LMC (Dufour 1984) and are both likely to support for the retarded star formation.

An alternative idea to explain starless GMC is that the GMCs in the LMC are of very recent formation, a situation possibly similar to the Maddalena's clouds and ON-1 cloud complex. It is well known that the LMC has a number of supershells expanding to accumulate the interstellar matter (Meaburn 1980; Oey 1996; Kim et al.\ 1999). Yamaguchi et al.\ (2001a) investigated the possible correlation between GMCs and supergiant shells and conclude that one third of GMCs may be located towards the shell boundaries, suggesting that a significant number of GMCs may have been formed under the triggering by expanding shells. The spatial distribution of the starless GMCs is however fairly random, showing little correlation with supergiant shells. It seems therefore the retarded star formation in the LMC is not due to some local environment or dynamical activities.

Finally, we shall comment on a possible link between Type I GMCs and formation of populous stellar clusters. It is tempting to speculate that there is a link between populous clusters and the Type I GMCs. A possible explanation is that a longer timescale of star formation allows the formation of proto-cluster molecular condensations as massive as $10^{5} M_{\sun}$, which can lead to form populous clusters. This will never happen in the Galaxy because of the star formation immediately after formation of proto-cluster condensations having mass of $10^{3} M_{\sun}$.

\subsection{Evolution of the GMCs}

The molecular clouds are considered to be formed in neutral gas, and as they evolve, they form stars and clusters, being dissipated by stellar winds or UV radiation from the massive stars or supernova explosions. At the end of their life, the newly formed stars and clusters remain. The process is, however, still unclear quantitatively, because we need a complete data set of molecular clouds to estimate the evolutionary time scale statistically. In our Galaxy, it is difficult to obtain such a complete sample due to the heavy contamination toward the Galactic plane. On the other hand, a face-on galaxy like the LMC is suitable for collecting a complete sample, which enables us to investigate the evolutionary process more quantitatively. By using the complete data set of the 
GMCs for a whole galaxy, we shall estimate their evolutionary time scale. In this section, only the GMCs with mass, $M_{\rm CO} > 5 \times 10^4 M_{\sun}$ are considered.

The evolutionary sequence of the GMCs is schematically drawn in Figure 13 together with examples of the GMCs correponding to each stage. In the first stage, the GMC Type I, the GMCs show no sign of massive star formation. The second stage, the GMC Type II, is the GMCs associated only with HII regions.  They are forming massive stars, but clusters have not appeared yet.  
The third, the GMC Type III, is the GMCs associated with both HII regions and clusters.  They are actively forming clusters. Molecular gas around newly formed clusters is partly dissipated, but the GMCs are still massive as is seen in Figure 12. The last is when the GMCs have been completely dissipated and only the young clusters and/or supernova remnants are found.

If we assume that the GMCs and clusters in the LMC are being formed nearly steadily, we can estimate each evolutionary time scale according to the above classification. First, we shall estimate the time scale of the cloud dissipation. We found that $\sim$ 66 \% of the youngest clusters of $\tau <$ 10 Myr are associated with the GMCs (see section 3.2). If we assume that the clusters in the LMC are being formed nearly steadily in the past 10 Myr, this result means the GMCs can survive during $\sim$ 60 \% of the cluster age, 10 Myr and are dissipated in a few Myr after formation of clusters due to the UV photons from the clusters.
The time scale for the GMC Type III is thus considered to be $\sim$ 7 Myr.  If we further assume that the massive star formation and cluster formation occur nearly steadily, the time scale for each stage is proportional to the number of the GMCs.  Accordingly, the time scale for the GMC Type I and II are estimated to be 6 Myr and 13 Myr, respectively. As a result, typical lifetime of the GMCs, corresponding to the total lifetime of the GMC from Type I to III, is roughly $\sim$ 30 Myr.  

Fukui et al. (1999) and Yamaguchi et al. (2001b) made a comparison of 55 GMCs with mass above $2 \times 10^5 M_{\sun}$ from the NANTEN first survey with the young objects. Their result shows that 6 GMCs show no sign of massive star formation, 9 are associated with HII regions, and 28 with young clusters, indicating that the population of Type III GMCs is the highest, 50 \%, and Type II and I are 21 \% and 14 \%, respectively. The higher sensitivity of the second survey made us possible to obtain a sample of the GMCs with mass as small as $5 \times 10^4 M_{\sun}$. This provides us with a complete sample of the GMCs, which are massive enough to produce massive stars. The numbers of Type I and Type II GMCs are increased in the current work compare to the 1st survey more than that of the Type III, because the mass of the Type I and Type II are smaller than the mass of the Type III, and thus, the higher sensitivity of the survey increased the number of Type I and Type II more. As a consequence of it, the estimated time scale of the a GMC now becomes longer from the previous result.                                                                                                                                              

We summarize the evolutionary time scale of the GMCs in Table 3.  The GMCs form massive stars, and HII regions appear after $\sim$ 6 Myr from their birth.  After 13 Myr, clusters start to be formed, then also start dissipating the surrounding molecular gas.  The GMCs continue to form clusters actively, being dissipated by the UV photons and stellar winds from the clusters.  After $\sim$ 7 Myr, the GMCs have been almost dissipated by the newly formed clusters, and eventually, by supernova explosions. The above time estimation should include an error. The age determination for the clusters contains uncertainty, but this changes only the absolute time scale for each stage. Other errors are possible due to a simple assumption of the constant formation rate for the clusters and the GMCs. Our estimation, nevertheless, shows a typical evolutionary sequence of the GMCs.  Further quantitative detailed studies, such as a comparison with H$\alpha$ flux, determining the age and mass of the clusters, and observations at higher resolution, will lead us to better understandings of star formation processes and the cloud dissipation in the LMC. Furthermore, detailed comparisons of the molecular clouds with HI give us a clue to understand the molecular clouds formation. These studies are found in elsewhere (e.g., Wong et al. 2009; Fukui et al. 2009)

\section{Summary}
We summarize our results obtained by comparing the molecular clouds from the second NANTEN survey (Paper I) with the young stellar clusters and HII regions.\\

\begin{enumerate}

\item We made a positional comparison of the molecular clouds with classical HII regions and clusters. It is indicated that the youngest group of the clusters, SWB 0 type, with an age of $\tau < 10$ Myr and HII regions show a significant correlation with the GMCs, while the clusters older than 10 Myr show little or no correlation.

\item The molecular clouds are classified into three Types; Type I shows no signature of star formation, Type II is associated with relatively small HII region(s) and Type III with both HII region(s) and young stellar cluster(s). Out of 272 molecular clouds, 72, 142 and 58 are found to be Type I, II, and III, respectively. The radio continuum sources were used to confirm that Type I molecular clouds do not host optically hidden HII regions. 

\item It is found that there is not a significant difference in the distribution of the line widths and sizes of the GMCs among the three Types for those with a mass above the completeness limit, $5 \times 10^{4} M_\sun$, while the mass distribution of the Type III GMCs is different from those of Type I and II. The mass distribution of Type I and II shows a peak at $M_{\rm CO} \sim 10^5 M_{\sun}$, while that of the Type III is rather flat.

\item We interpret that these Types represent the evolutionary sequence; i.e., the youngest phase is Type I followed by Type II and the last phase is Type III where most active star formation takes place leading to cloud dispersal. The number of the three types of GMCs should be proportional to the time scale of each evolutionary stage if a steady state is a good approximation. By adopting the time scale of the youngest stellar clusters, 10 Myrs, we roughly estimate the timescales of Types I, II and III to be 6 Myrs, 13 Myrs and 7 Myrs, respectively, for those with a mass above the completeness limit, $5 \times 10^{4} M_\sun$. This corresponds to a lifetime of the GMC of 20--30 Myrs.
\end{enumerate}

\acknowledgments
The NANTEN project is based on a mutual agreement between Nagoya University and the Carnegie Institution of Washington (CIW). We greatly appreciate the hospitality of all the staff members of the Las Campanas Observatory of CIW. 
We are thankful to many Japanese public donors and companies who contributed to the realization of the project. This study has made use of SIMBAD Astronomical Database and NASA's Astrophysics Data System Bibliographic Services. This work is financially supported in part by a Grant-in-Aid for Scientific Research from the Ministry of Education, Culture, Sports, Science and Technology of Japan (No. 15071203), from JSPS (No. 14102003, No. 18684003, and core-to-core program
17004), and the Mitsubishi Foundation.
AK is financially supported by the 21st century COE program.

\clearpage
\begin{deluxetable}{llcc}
\tablecolumns{4} 
\tablewidth{0pc} 
\tabletypesize{\footnotesize}
\tablecaption{Number of clusters and OB associations} 
\tablehead{\colhead{SWB type\tablenotemark{a}} &
\colhead{age\tablenotemark{a}} &
\colhead{Number of clusters\tablenotemark{a,b}}&
\colhead{Number of clusters\tablenotemark{b} in observed area} \\
 &\colhead{(Myr)} & &}
\startdata
SWB 0 & \phn\phn\phn\phn\phn$<$ 10  & 137 & 137 \\
SWB I & \phn\phn10\phn--\phn30 & 130 & 122 \\
SWB II & \phn\phn30\phn--\phn70 & \phn65 & \phn56  \\
SWB III & \phn\phn70\phn--\phn200 & \phn87 & \phn71  \\
SWB IVA & \phn200\phn--\phn400 & \phn62 & \phn45 \\
SWB IVB & \phn400\phn--\phn800 & \phn33 & \phn23 \\
SWB V & \phn800\phn--\phn2000 & \phn41 & \phn16 \\
SWB VI & 2000\phn--\phn5000 & \phn30 & \phn11  \\
SWB VII & 5000\phn--\phn16000 & \phn38 & \phn17  \\
\enddata
\tablenotetext{a}{from Bica et al.\ (1996).}
\tablenotetext{b}{including both clusters and OB associations.}
\end{deluxetable}
\begin{deluxetable}{cclllcc}
\rotate
\tablewidth{480pt}
\setlength{\tabcolsep}{0.04in}
\tablecolumns{7} 
\tablewidth{0pc} 
\tabletypesize{\footnotesize}
\tablecaption{Associated HII regions, OB associations, and young clusters} 
\tablehead{ \multicolumn{2}{c}{Molecular clouds\tablenotemark{a}}&
\multicolumn{3}{c}{Young Objects}&\colhead{GMC Type}& \colhead{Notes}\\
\cline{3-5}\\
 & & \multicolumn{2}{c}{HII regions\tablenotemark{b}} & 
 \colhead{Young clusters\tablenotemark{c}} & &} 
\startdata
1   &   LMC N J0447-6910   &      &      &      &   I   &   \\
2   &   LMC N J0447-6713   &      &      &      &   I   &   1   \\
3   &   LMC N J0448-6920   &   N 79   &   DEM 6   &      &   II   &   \\
4   &   LMC N J0449-6910   &   N 77   &   DEM 2, DEM 4, DEM 5   &   IC 2105   &   III   &   \\
5   &   LMC N J0449-6826   &   N 76   &   DEM 3   &      &   II   &   \\
6   &   LMC N J0449-6806   &      &      &      &   I   &   \\
7   &   LMC N J0449-6652   &      &      &      &   I   &   \\
8   &   LMC N J0450-6910   &  N 77  &   DEM 4   &      &   II   &   \\
9   &   LMC N J0450-6930   &   N 78, N 79   &   DEM 6   &      &   II   &   \\
10   &   LMC N J0450-6919   &   N 79   &   DEM 6, DEM 10   &      &   II   &   \\
11   &   LMC N J0451-6858   &      &      &      &   I   &   \\
12   &   LMC N J0451-6704   &   N 4, N 5   &   DEM 8, DEM 11, DEM 12   &   NGC 1714, SL  69   &   III   &   \\
&  &   & DEM 18   &   &   &   \\
13   &   LMC N J0451-6729   &      &      &      &   I   &   \\
14   &   LMC N J0451-6922   &   N 79   &   DEM 10, DEM 15   &  BRHT 1a,  IC 2111, NGC 1727 &   III   &   \\
15   &   LMC N J0452-6750   &      &      &      &   I   &   \\
16   &   LMC N J0452-6804   &   N 8   &   DEM 13   &   NGC 1736   &   III   &   \\
17   &   LMC N J0453-6909   &   N 81, N 83, N 90   &   DEM 15, DEM 22, DEM 36   &   NGC 1737, NGC 1743 &   III   &   \\
   &      &   &   & NGC 1745, NGC 1748   &    &   \\
18   &   LMC N J0453-6919   &   N 79, N 82, N 88, N 89   &   DEM 10, DEM 15   &      &   II   &   \\
19   &   LMC N J0454-6644   &   N 6   &   DEM 27, DEM 29   &      &   II   &   \\
20   &   LMC N J0454-6755   &      &      &      &   I   &   \\
21   &   LMC N J0454-6929   &   N 87   &   DEM 23   &      &   II   &   \\
22   &   LMC N J0455-6830   &   N 80, N 84, N 85, N 86   &   DEM 26, DEM 32, DEM 33   &      &   II   &   \\
23   &   LMC N J0455-6634   &   N 11   &   DEM 34   &   NGC 1760, NGC 1761  &   III   &   \\
24   &   LMC N J0455-6930   &   N 94   &   DEM 36   &   SL 109   &   III   &   \\
25   &   LMC N J0457-6910   &   N 93   &   DEM 37   &      &   II   &   \\
26   &   LMC N J0457-6844   &   N 92   &   DEM 38   &   LH 11   &   III   &   \\
27   &   LMC N J0457-6826   &   N 91   &   DEM 39   &   KMHK341,KMHK345   &   III   &   \\
28   &   LMC N J0457-6856   &      &      &      &   I   &   \\
29   &   LMC N J0457-6702   &      &    &   &   I   &   \\
30   &   LMC N J0457-6620   &   N 11   &   DEM 34   & IC 2116, NGC 1763e &   III   &   \\
   &     &      &   &   NGC 1763w   &     &   \\
31   &   LMC N J0457-6628   &   N 11   &   DEM 34   & HNT 1, NGC 1769   &   III   &   \\
32   &   LMC N J0458-6617   &   N 11   &   DEM 34, DEM 41   &      &   II   &   \\
33   &   LMC N J0458-6618   &   N 11   &   DEM 34, DEM 41   &      &   II   &   \\
34   &   LMC N J0458-7022   &      &      &      &   I   &  2 \\
35   &   LMC N J0459-6614   &   N 11, N12   &   DEM 34, DEM 42, DEM 44,&      &   II   &   \\
   &   &   &   DEM 46, DEM 47   &      &     &   \\
36   &   LMC N J0500-6622   &   N 15   &   DEM 49   &      &   II   &   \\
37   &   LMC N J0501-6804   &     &    &      &  I   &   \\
38   &   LMC N J0502-6903   &      &   DEM 60, DEM 62   &      &   II   &   \\
39   &   LMC N J0503-6553   &      &   DEM 61   &      &   II   &   \\
40   &   LMC N J0503-6540   &      &   DEM 48   &      &   II   &   \\
41   &   LMC N J0503-6828   &   N 95   &   DEM 58   &      &   II   &   \\
42   &   LMC N J0503-6612   &      &      &      &   I   &   \\
43   &   LMC N J0503-6643   &      &   DEM 56   &      &   II   &   \\
44   &   LMC N J0503-6711   &   N 17   &   DEM 59   &   &   II   &   \\
45   &   LMC N J0503-6719   &   N 17   &   DEM 59   &   NGC 1814, NGC 1816  &   III   &   \\
   &   &    &   &  NGC 1820   &  &   \\
46   &   LMC N J0504-6802   &   N 23   &   DEM 66, DEM 70   &   NGC 1829   &   III   &   \\
47   &   LMC N J0504-6639   &   N 18   &   DEM 56   &      &   II   &   \\
48   &   LMC N J0504-7007   &   N 189   &   DEM 67, DEM 68   &      &   II   &   \\
49   &   LMC N J0504-7056   &   N 191   &   DEM 64, DEM 75   &   LH 23, N 191A   &   III   &   \\
50   &   LMC N J0505-6650   &   N 20   &   DEM 69   &      &   II   &   \\
51   &   LMC N J0506-7010   &   N 189   &   DEM 68   &      &   II   &   \\
52   &   LMC N J0506-6753   &      &   DEM 78   &      &   II   &   \\
53   &   LMC N J0507-7041   &      &   DEM 80   &      &   II   &   \\
54   &   LMC N J0507-6828   &   N 100   &   DEM 76, DEM 79   &      &   II   &   \\
55   &   LMC N J0507-6858   &      &      &      &   I   &   \\
56   &   LMC N J0508-6902   &      &      &      &   I   &   \\
57   &   LMC N J0508-6905   &      &      &      &   I   &   \\
58   &   LMC N J0508-6923   &   N 108   &   DEM 92   &      &   II   &   \\
59   &   LMC N J0509-6844   &   N 103   &   DEM 85   &   NGC 1850A   &   III   &   \\
60   &   LMC N J0509-6827   &   N 104   &   DEM 88   &      &   II   &   \\
61   &   LMC N J0509-7049   &   N 192   &   DEM 80   &      &   II   &   \\
62   &   LMC N J0509-6912   &      &      &      &   I   &   \\
63   &   LMC N J0510-6853   &   N 105, N 106   &   DEM 86, DEM 87   &   NGC 1858   &   III   &   \\
64   &   LMC N J0510-6706   &   N 26, N 27   &   DEM 90, DEM 93, DEM 94    &      &   II   &   \\
65   &   LMC N J0511-6927   &   N 108, N109   &   DEM 92   &      &   II   &   \\
66   &   LMC N J0512-6746   &      &      &      &   I   &   \\
67   &   LMC N J0512-6710   &      &   DEM 97, DEM 98   &      &   II   &   \\
68   &   LMC N J0512-6903   &   N 111   &   DEM 95, DEM 96, DEM 103   &      &   II   &   \\
69   &   LMC N J0512-7028   &   N 192, N 193   &   DEM 100, DEM 101, DEM 102   &      &   II   &   \\
70   &   LMC N J0512-6811   &      &      &      &   I   &   \\
71   &   LMC N J0513-6936   &   N 109, N 114   &   DEM 110   &      &   II   &   \\
72   &   LMC N J0513-6922   &   N 113, N 114   &   DEM 104, DEM 108, DEM 113   &   NGC 1874, NGC 1876, SL 320   &   III   &   \\
73   &   LMC N J0513-6723   &   N 30   &   DEM 105   &   NGC 1873   &   III   &   \\
74   &   LMC N J0514-7010   &   N 115   &   DEM 114   &      &   II   &   \\
75   &   LMC N J0514-6847   &      &      &      &   I   &   \\
76   &   LMC N J0514-6727   &   N 30   &   DEM 106   &      &   II   &   \\
77   &   LMC N J0515-7034   &      &      &      &   I   &   \\
78   &   LMC N J0515-7002   &      &      &      &   I   &   \\
79   &   LMC N J0515-6857   &      &      &      &   I   &   \\
80   &   LMC N J0516-6807   &   N 32   &      &      &   II   &   \\
81   &   LMC N J0516-6616   &      &      &      &   I   &   \\
82   &   LMC N J0516-6847   &      &      &      &   I   &   \\
83   &   LMC N J0516-6922   &      &   DEM 130   &      &   II   &   \\
84   &   LMC N J0516-6559   &   N 35   &   DEM 125, DEM 126   &      &   II   &   \\
85   &   LMC N J0517-7046   &      &      &      &   I   &   \\
86   &   LMC N J0517-7114   &   N 195   &   DEM 131   &   NGC 1914   &   III   &   \\
87   &   LMC N J0517-6847   &   N 119   &   DEM 128   &      &   II   &   \\
88   &   LMC N J0517-6915   &   N 119   &   DEM 123, DEM 130, DEM 132   &   H88-266, NGC1910, SL 360   &   III   &   \\
89   &   LMC N J0517-6642   &   N 34   &      &      &   II   &   \\
90   &   LMC N J0517-6932   &      &   DEM 130, DEM 134   &      &   II   &   \\
91   &   LMC N J0517-6922   &   N 119   &   DEM 130, DEM 132   &      &   II   &   \\
92   &   LMC N J0518-7001   &      &      &      &   I   &   \\
93   &   LMC N J0518-6620   &      &      &      &   I   &   \\
94   &   LMC N J0518-6951   &   N 121, N 123   &   DEM 133   &      &   II   &   \\
95   &   LMC N J0519-6625   &      &      &      &   I   &   \\
96   &   LMC N J0519-6938   &  N 120, N 121, N 122   &   DEM 133, DEM 134   &   NGC 1918   &   III   &   \\
97   &   LMC N J0519-7113   &      &      &      &   I   &   \\
98   &   LMC N J0519-6908   &   N 119   &   DEM 123, DEM 132   &  &   II   &   \\
99   &   LMC N J0520-7043   &      &      &      &   I   &   \\
100   &   LMC N J0520-6651   &   N 37, N 38   &   DEM 136, DEM 138   &      &   II   &   \\
101   &   LMC N J0520-6837   &   N 124, N 128   &   DEM 157   &      &   II   &   \\
102   &   LMC N J0520-6832   &      &      &      &   I   &   \\
103   &   LMC N J0521-7143   &   N 197   &   DEM 146   &      &   II   &   \\
104   &   LMC N J0521-7013   &   N 125, N 130   &   DEM 163   &      &   II   &   \\
105   &   LMC N J0521-7000   &   N 123, N 125, N 130   &   DEM 147, DEM 148   &      &   II   &   \\
106   &   LMC N J0521-6841   &   N 124, N 128   &   DEM 157   &      &   II   &   \\
107   &   LMC N J0521-6801   &   N 41, N 44   &   DEM 140, DEM 141, DEM 152 &   N44K, NGC 1936   &   III   &   \\
   &    &   & DEM 158   &   &     &   \\
108   &   LMC N J0521-6847   &   N 126   &   DEM 143   &      &   II   &   \\
109   &   LMC N J0521-6948   &      &      &      &   I   &   \\
110   &   LMC N J0522-6941   &   N 127, N 129   &   DEM 145, DEM 149,  DEM 153   &   HS248   &   III   &   \\
111   &   LMC N J0522-7135   &   N 198   &   DEM 165   &      &   II   &   \\
112   &   LMC N J0522-6828   &      &      &      &   I   &   \\
113   &   LMC N J0522-6833   &  N 128 &   DEM 157   &      &   II   &   \\
114   &   LMC N J0522-6756   &   N 41, N 44  & DEM 144, DEM 150, DEM 151 & BCDSP 2, N44K, NGC 1929 &   III   &   \\
      &                      &               & DEM 152, DEM 156, DEM 159 & NGC 1934, NGC 1935, NGC 1936 &    &   \\
      &                      &               & DEM 160, DEM 166, DEM 167 & NGC 1937, SL 417P1, SL 429 &    &   \\
      &                      &               & DEM 169, DEM 170          &   &    &   \\
115   &   LMC N J0522-6541   &   N 43   &   DEM 142, DEM 155 &   SL 428   &   III   &   \\
116   &   LMC N J0522-6951   &   N 131   &   DEM 168   &      &   II   &   \\
117   &   LMC N J0522-6706   &      &      &      &   I   &  3  \\
118   &   LMC N J0523-6820   &      &      &      &   I   &   \\
119   &   LMC N J0523-6642   &   N 45   &   DEM 161   &      &   II   &   \\
120   &   LMC N J0523-6629   &      &      &      &   I   &   \\
121   &   LMC N J0523-6706   &      &      &      &   I   &  3   \\
122   &   LMC N J0523-6706   &      &      &      &   I   &   3   \\
123   &   LMC N J0523-7139   &   N 198   &   DEM 165, DEM 194   &      &   II   &   \\
124   &   LMC N J0523-6902   &   N 136   &   DEM 177   &      &   II   &   \\
125   &   LMC N J0524-6828   &   N 138   &   DEM 174, DEM 179, DEM 180 &   NGC 1949   &   III   &   \\
126   &   LMC N J0524-6727   &   N 51   &   DEM 192   &      &   II   &   \\
127   &   LMC N J0524-7027   &      &      &      &   I   & 4 \\
128   &   LMC N J0524-6826   &   N 138   &   DEM 174, DEM 179, DEM 180 &   NGC 1949   &   III   &   \\
 &     &   &   DEM 203   &     &   &   \\
129   &   LMC N J0524-6939   &  N 132  &   DEM 172, DEM 173, DEM 186   &      &   II   &   \\
130   &   LMC N J0524-6914   &   N 139, N 140   &   DEM 177   &      &   II   &   \\
131   &   LMC N J0524-7134   &   N 201, N 205   &   DEM 194, DEM 202   &      &   II   &   \\
132   &   LMC N J0525-6948   &   N 134   &      &      &   II   &   \\
133   &   LMC N J0525-6918   &   N 139, N 140, N 142, &   DEM 177, DEM 187, DEM 197   &   SL 475   &   III   &   \\
 &   & N 143   & &  &  &   \\
134   &   LMC N J0525-6622   &   N 48   &      &      &   II   &   \\
135   &   LMC N J0525-6609   &   N 48, N 49 &   DEM 175, DEM 181, DEM 185 &   NGC  1948, BSDL 1610   &   III   &   \\
   &   &  &   DEM 189, DEM 190, DEM 191   &   &      &   \\
   &   &  &   DEM 195   &   &      &   \\
136   &   LMC N J0526-6835   &      &   DEM 203   &      &   II   &   \\
137   &   LMC N J0526-6839   &      &   DEM 203   &      &   II   &   \\
138   &   LMC N J0526-6742   &   N 51   &   DEM 192, DEM 196, DEM 211   &   SL 471   &   III   &   \\
139   &   LMC N J0526-6848   &   N 144   &   DEM 199   &   NGC 1962, NGC 1965   &   III   &   \\
   &    &  &   &   NGC 1966   &     &   \\
140   &   LMC N J0526-6748   &      &   DEM 196   &      &   II   &   \\
141   &   LMC N J0526-6552   &      &   DEM 185, DEM 204   &      &   II   &   \\
142   &   LMC N J0526-6853   &   N 141, N 144   &   DEM 199   &      &   II   &   \\
143   &   LMC N J0526-7119   &   N 205   &   DEM 206, DEM 207   &      &   II   &   \\
144   &   LMC N J0527-7028   &   N 204   &   DEM 208   &      &   II   &   \\
145   &   LMC N J0527-7035   &   N 204   &   DEM 208   &      &   II   &   \\
146   &   LMC N J0527-7053   &   N 206   &   DEM 221   &      &   II   &   \\
147   &   LMC N J0528-7140   &      &      &      &   I   &   \\
148   &   LMC N J0528-6953   &      &   DEM 215   &      &   II   &   \\
149   &   LMC N J0528-7058   &   N 206   &   DEM 221   &      &   II   &   \\
150   &   LMC N J0529-6830   & N 148 &   DEM 203  &      &   II   &   \\
151   &   LMC N J0530-7128   &      &      &      &   I   &   \\
152   &   LMC N J0530-7054   &   N 206   &   DEM 221   &      &   II   &   \\
153   &   LMC N J0530-7106   &   N 206   &   DEM 221   &   LH 69P1, NGC 2018   &   III   &   \\
154   &   LMC N J0531-6830   &   N 148  &   DEM 203, DEM 226&   CBDSP 7, LH 71   &   III   &   \\
155   &   LMC N J0532-6742   &   N 57   &   DEM 219, DEM 229, DEM 230 &   BCDSP 6, NGC 2014   &   III   &   \\
  &   &    &   DEM 231   &   &     &   \\
156   &   LMC N J0532-7114   &   N 206   &   DEM 221   &      &   II   &   \\
157   &   LMC N J0532-6838   &   N 148   &   DEM 226   &    &   II   &   \\
158   &   LMC N J0532-6626   &   N 55   &   DEM 227, DEM 228   &  SL 553n, SL 553s    &   III   &   \\
159   &   LMC N J0532-6946   &   N 149   &   DEM 246   &      &   II   &   \\
160   &   LMC N J0532-6854   &      &   DEM 232   &    &   II   &   \\
161   &   LMC N J0532-6730   &   N 58   &   DEM 234   &   SL 567   &   III   &   \\
162   &   LMC N J0532-6857   &      &   DEM 232   &      &   II   &   \\
163   &   LMC N J0533-6819   &   N 148   &   DEM 226  &      &   II   &   \\
164   &   LMC N J0534-6813   &      &      &      &   I   &   \\
165   &   LMC N J0535-6840   &      &   DEM 232   &      &   II   &   \\
166   &   LMC N J0535-6742   &      &   DEM 245   &      &   II   &   \\
167   &   LMC N J0535-7045   &      &      &      &   I   &   \\
168   &   LMC N J0535-6735   &   N 59   &  DEM 241    &   NGC 2029, NGC 2032&   III   &   \\
  &   &   &      & NGC 2035, NGC 2040   & &   \\
169   &   LMC N J0535-6912   &   N 157   &   DEM 263   &   BRHT 17b   &   III   &   \\
170   &   LMC N J0535-6912   &   N 157   &   DEM 263   &   BRHT 17b   &   III   &   \\
171   &   LMC N J0535-6902   &  N 153, N 157   &   DEM 263   &   LH 89s   &   III   &   \\
172   &   LMC N J0535-6844   &      &   DEM 232   &      &   II   &   \\
173   &   LMC N J0536-7043   &      &      &      &   I   &   \\
174   &   LMC N J0536-6931   &  N 154  &   DEM 248, DEM 262   &      &   II   &   \\
175   &   LMC N J0536-6824   &      &   DEM 268   &      &   II   &   \\
176   &   LMC N J0536-7124   &      &      &      &   I   &   \\
177   &   LMC N J0537-6651   &      &   DEM 257   &      &   II   &   \\
178   &   LMC N J0537-6946   &   N 155   &   DEM 246, DEM 248, DEM 259 &      &   II   &   \\
  &   &    &   DEM 260, DEM 262   &      &    &   \\
179   &   LMC N J0537-6619   &   N 64   &   DEM 252, DEM 253, DEM 254   &   LH 95,HS367   &   III   &   \\
180   &   LMC N J0537-6626   &      &   DEM 251, DEM 256   &      &   II   &   \\
181   &   LMC N J0537-6944   &   N 155   &   DEM 246, DEM 248, DEM 259 &      &   II   &   \\
    &    &   &  DEM 260, DEM 262   &      &     &   \\
182   &   LMC N J0537-6946   &   N 155   &   DEM 260   &      &   II   &   \\
183   &   LMC N J0538-6908   &   N 157   &   DEM 263   &   30 Dor, NGC 2060   &   III   &   \\
184   &   LMC N J0538-6934   &   N 158   &   DEM 263, DEM 269   &      &   II   &   \\
185   &   LMC N J0538-7117   &      &      &      &   I   &   \\
186   &   LMC N J0538-6904   &   N 157   &   DEM 263   &   30 Dor   &   III   &   \\
187   &   LMC N J0538-7119   &      &      &      &   I   &   \\
188   &   LMC N J0538-7017   &   N 171   &   DEM 267   &      &   II   &   \\
189   &   LMC N J0538-6915   &   N 157   &   DEM 263, DEM 269   &   NGC 2060  &   III   &   \\
190   &   LMC N J0538-6854   &   N 161   &   DEM 263 &  NGC 2069    &   III   &   \\
191   &   LMC N J0539-6932   &  N 158  &   DEM 263, DEM 269   &  NGC 2074  &   III   &   \\
192   &   LMC N J0539-6954   &   N 172   &   DEM 275   &      &   II   &   \\
193   &   LMC N J0539-7017   &   N 171   &   DEM 267   &      &   II   &   \\
194   &   LMC N J0539-6918   &  N 158  &   DEM 263, DEM 269   &      &   II   &   \\
195   &   LMC N J0539-7109   &   N 214   &   DEM 274, DEM 276, DEM 278  &      &   II   &   \\
   &  &  &    DEM 287, DEM 288   &      &  &   \\
196   &   LMC N J0539-7132   &      &      &      &   I   &   \\
197   &   LMC N J0540-7008   &   N 158, N 159, N 160&   DEM 260, DEM 265, DEM 267 &   LH 103PI, NGC 2077 &   III   &   \\
      &      &N 171, N 172, N 173, &  DEM 269, DEM 271, DEM 272 &   NGC 2078, NGC 2079  &     &   \\
      &      &N 174, N 175, N 176, &DEM 275, DEM 277,  DEM 279&   NGC 2083, NGC 2084e &      &   \\
        &      & N 177, N 178, N 213 &DEM 280, DEM 281, DEM 282, &   NGC 2084w, NGC 2084 &    &   \\
        &      & N 216, N 217, N 218 &  DEM 291, DEM 294, DEM 295  &   NGC 2085, NGC 2086   &  &   \\
198   &   LMC N J0540-7113   &   N 214   &   DEM 288   &      &   II   &   \\
199   &   LMC N J0540-7124   &      &   DEM 293   &      &   II   &   \\
200   &   LMC N J0540-6905   &   N 161   &   DEM 263, DEM 273   &      &   II   &   \\
201   &   LMC N J0540-6900   &   N 161   &   DEM 263, DEM 273   &      &   II   &   \\
202   &   LMC N J0540-7003   &   N 175   &   DEM 281, DEM 295   &      &   II   &   \\
203   &   LMC N J0541-7053   &   N 216   &   DEM 290   &      &   II   &   \\
204   &   LMC N J0541-6900   &      &   DEM 263, DEM 297   &      &   II   &   \\
205   &   LMC N J0542-7119   &   N 214   &   DEM 288, DEM 289, DEM 292  &   NGC 2103   &   III   &   \\
   &   &   &   DEM 293   &  &   &   \\
206   &   LMC N J0542-6943   &   N 163   &   DEM 300   &      &   II   &   \\
207   &   LMC N J0542-6944   &   N 163   &   DEM 300   &      &   II   &   \\
208   &   LMC N J0543-6757   &   N 70   &   DEM 301   &   LH 114   &   III   &   \\
209   &   LMC N J0543-6926   &  N 166  &   DEM 304   &      &   II   &   \\
210   &   LMC N J0543-7012   &      &      &      &   I   &   \\
211   &   LMC N J0543-6754   &   N 70   &   DEM 301   &   LH 114   &   III   &   \\
212   &   LMC N J0543-7106   &      &      &      &   I   &   \\
213   &   LMC N J0543-6726   &   N 71, N 73   &   DEM 303, DEM 305   &   N 71   &   III   &   \\
214   &   LMC N J0543-6915   &  N 167  &   DEM 304, DEM 307, DEM 310   &      &   II   &   \\
215   &   LMC N J0544-7127   &      &      &      &   I   &   \\
216   &   LMC N J0544-6923   & N 166, N 167  &   DEM 310   &      &   II   &   \\
217   &   LMC N J0544-6943   &   N 168   &   DEM 311   &      &   II   &   \\
218   &   LMC N J0544-6718   &      &   DEM 308   &      &   II   &   \\
219   &   LMC N J0545-6856   &      &      &      &   I   &   \\
220   &   LMC N J0545-6949   &   N 168   &   DEM 311   &  NGC 2113    &   III   &   \\
221   &   LMC N J0545-6708   &   N 74   &   DEM 309  &      &   II   &   \\
222   &   LMC N J0546-7105   &      &      &      &   I   &   \\
223   &   LMC N J0546-6937   &   N 169   &   DEM 312, DEM 313, DEM 314  &      &   II   &   \\
224   &   LMC N J0547-6808   &      &      &      &   I   &   \\
225   &   LMC N J0547-7041   &      &      &      &   I   &   \\
226   &   LMC N J0547-6953   &   N 179   &   DEM 316, DEM 319, DEM 320, DEM 321   &      &   II   &   \\
   &      &     &  DEM 324   &      &    &   \\
227   &   LMC N J0548-7007   &   N 180   &   DEM 317, DEM 322, DEM 323, DEM 325   &   NGC 2122   &   III   &   \\
   &   &   &DEM 326   &   &    &   \\
228   &   LMC N J0548-7034   &      &      &      &   I   &   \\
229   &   LMC N J0550-6827   &      &      &      &   I   &   \\
230   &   LMC N J0555-6811   &   N 75   &      &   SL 785   &   III   &   \\
231   &   LMC N J0447-6728   &      &      &      &   I   &   \\
232   &   LMC N J0448-6721   &      &      &      &   I   &   \\
233   &   LMC N J0449-6810   &      &     &      &   I   &   \\
234   &   LMC N J0451-6927   &   N 79   &   DEM 6, DEM 10   &      &   II   &   \\
235   &   LMC N J0453-6846   &      &   DEM 16   &      &   II   &   \\
236   &   LMC N J0458-6617   & N 11  &   DEM 34   &      &   II   &   \\
237   &   LMC N J0459-6607   &   N 13   &   DEM 47   &      &   II   &   \\
238   &   LMC N J0500-6807   &   N 16   &   DEM 45, DEM 52   &      &   II   &   \\
239   &   LMC N J0504-6904   &    &   DEM 62   &      &   II   &   \\
240   &   LMC N J0505-6950   &     &      &      &   I   &   \\
241   &   LMC N J0507-6647   &      &      &      &   I   &   \\
242   &   LMC N J0511-6815   &      &      &      &   I   &   \\
243   &   LMC N J0511-7050   &      &      &      &   I   &   \\
244   &   LMC N J0511-6928   &   N 109   &      &      &   II   &   \\
245   &   LMC N J0511-6857   &      &   DEM 103   &      &   II   &   \\
246   &   LMC N J0521-6600   &      &   DEM 154   &      &   II   &   \\
247   &   LMC N J0521-6733   &      &      &      &   I   &   \\
248   &   LMC N J0523-7029   &      &      &      &   I   &   \\
249   &   LMC N J0525-7135   &   N 205   &   DEM 202   &      &   II   &   \\
250   &   LMC N J0525-6555   &      &   DEM 185   &      &   II   &   \\
251   &   LMC N J0526-6733   &   N 51   &   DEM 205   &      &   II   &   \\
252   &   LMC N J0528-6725   &  N 51  &   DEM 205   &   NGC 1974   &   III   &   \\
253   &   LMC N J0530-6757   &      &      &      &   I   &   \\
254   &   LMC N J0532-7056   &   N 206   &   DEM 221   &      &   II   &   \\
255   &   LMC N J0532-6735   &  N 57 &   DEM 229   &      &   II   &   \\
256   &   LMC N J0534-6823   &   N 148   &   DEM 226   &      &   II   &   \\
257   &   LMC N J0535-6730   & N 59 &  DEM 241    &      &   II   &   \\
258   &   LMC N J0535-6618   &   N 62   &   DEM 239   &      &   II   &   \\
259   &   LMC N J0535-6714   &   N 66   &   DEM 257   &    &   II   &   \\
260   &   LMC N J0536-6941   &   N 154   &   DEM 246   &   LH 87s, NGC 2048  &   III   &   \\
261   &   LMC N J0536-6850   &   N 157   &   DEM 263   &      &   II   &   \\
262   &   LMC N J0538-7006   &     &  DEM 266   &      &   II   &   \\
263   &   LMC N J0538-7040   &   N 213   &   DEM 265   & NGC 2075   &   III   &   \\
264   &   LMC N J0540-7100   &      &      &      &   I   &   \\
265   &   LMC N J0541-6815   &      &      &      &   I   &   \\
266   &   LMC N J0543-6947   &   N 163   &   DEM 300   &      &   II   &   \\
267   &   LMC N J0543-6749   &   N 70   &   DEM 301   &   LH 114   &   III   &   \\
268   &   LMC N J0543-6920   &  N 167  &   DEM 304   &      &   II   &   \\
269   &   LMC N J0543-6618   &   N 72   &   DEM 302   &      &   II   &   \\
270   &   LMC N J0547-6959   &   N 180   &   DEM 322   &      &   II   &   \\
271   &   LMC N J0553-6822   &      &      &      &   I   &   \\
272   &   LMC N J0554-6833    &      &      &      &   I   &   \\
\enddata
\tablecomments{1. H$\alpha$ source is seen at the peak of the cloud. 
2. H$\alpha$ source is seen within the cloud. 
3. Diffuse H $\alpha$ emission, a part of the LMC 4, is seen across the cloud. 
4. H$\alpha$ source, possibly an OB star (FAUST 840; UV emission source) is seen within the cloud.}
\tablenotetext{a}{Cloud identified in Paper I. The number and name of the clouds are from Tables 1 and 2 of Paper I.}
\tablenotetext{b}{HII regions associated with the clouds; "N" for Henize (1956), and "DEM" for Davies et al.\ (1976).}
\tablenotetext{c}{SWB0 clusters and associations by Bica et al.\ (1996)}

\end{deluxetable}
\begin{deluxetable}{llcccl}
\rotate
\tablecolumns{6} 
\tablewidth{0pc} 
\tabletypesize{\footnotesize}
\tablecaption{GMC Type and evolution of the GMCs} 
\tablehead{GMC &Observed signature & \multicolumn{2}{c}{Number of clouds} & Time scale \tablenotemark{a} & Star formation activities\\
\cline{3-4}
Type & &all clouds & GMCs\tablenotemark{a} & & \\
 & & & &(Myr) & }
\startdata
I\dotfill & no HII regions or young clusters & \phn72
 & 46(24\%) & \phn6 &  Without massive star formation \\
II\dotfill &with HII region(s) & 142 & 96(50\%) & 
13  &  Massive star formation \\
III\dotfill &with HII region(s) and SWB 0 cluster(s) & \phn58
& 49(26\%) & \phn7  & Cluster formation \\
\dotfill &young cluster only& \nodata & \nodata& \phn3 & Dissipation of clouds\\
\enddata
\tablenotetext{a}{Molecular clouds with $M_{\rm CO} > 5 \times 10^4M_\sun$.}
\tablecomments{Time scale of the GMCs at each evolutionary stage 
is estimated by assuming the formation of the
molecular clouds ($M_{\rm CO} > 5\times 10^{4} M_{\sun}$)
and clusters to be constant, and the time scale is
propotional to the number of the sample.}
\end{deluxetable}
\begin{deluxetable}{cccccccc}
\tablecolumns{11} 
\tablewidth{0pc} 
\tabletypesize{\footnotesize}
\tablecaption{Physical properties of the GMCs} 
\tablehead{GMC Type& Numer of clouds 
&  \multicolumn{2}{c}{Line width} 
& \multicolumn{2}{c}{Size} 
& \multicolumn{2}{c}{Mass}\\
\cline{3-4}\cline{5-6}\cline{7-8}\\
& 
&  $<\Delta V>$& $\sigma_{\Delta V}$ 
& $<R>$ & $\sigma_{R}$ 
& $<M_{CO}>$ & $\sigma_{M_{CO}}$   \\
& & (km s$^{-1}$) & (km s$^{-1}$)  
& (pc) & (pc) 
 &($10^5 M_\sun$) &($10^5 M_\sun$)}
\startdata
I\dotfill & 46(24\%) & 5.9 & 2.5 & 40 & 15 & 2 & \phn2 \\
II\dotfill &96(50\%) & 5.5 & 2.2 & 36 & 19 &  3 & \phn4 \\
III\dotfill & 49(26\%) & 7.0 & 3.0 & 46 & 30 & 6 & \phn20\\
\enddata
\tablecomments{Properties of the GMCs
with mass, $M_{\rm CO} > 5\times 10^{4} M_{\sun}$.
The average values of line width, size and mass for each Type
are shown in $< >$.
The average and $\sigma$ of the size are derived for 164 GMCs
with minor axis more than the NANTEN beam (see also Paper I).}
\end{deluxetable}

\clearpage


\begin{figure}
\epsscale{0.7}
\plotone{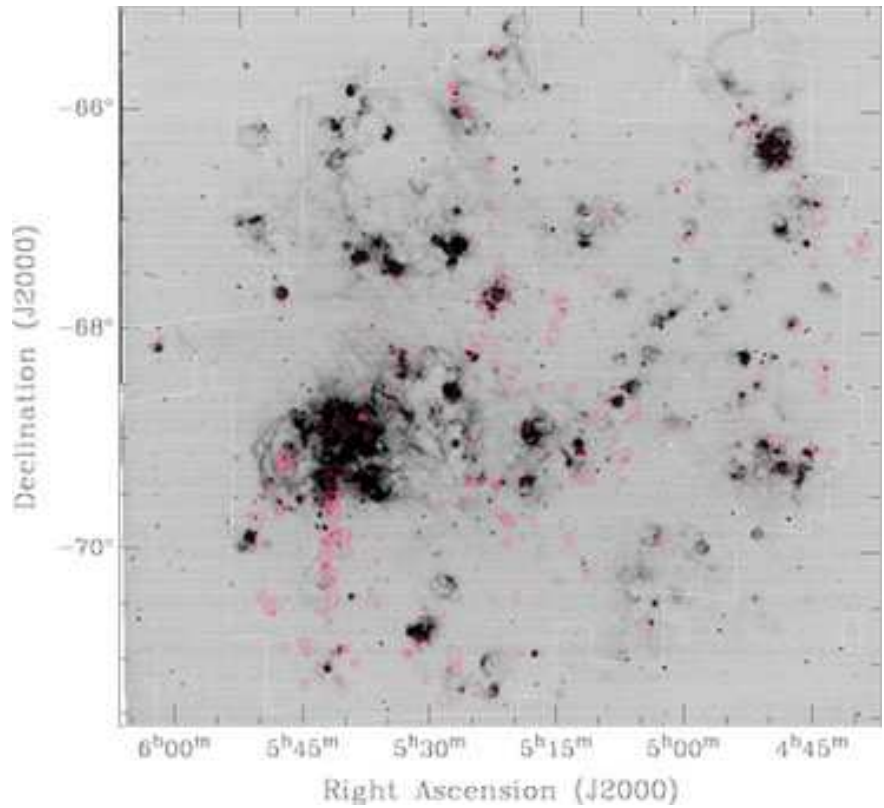}
\caption{
Distribution of the CO emission (Paper I) superposed on the H $\alpha$ image (Kim et al.\ 1999). The red contours show the velocity-integrated intensity of the CO (1--0) line obtained by NANTEN; the contours are from 1.2 K km s$^{-1}$ (3 $\sigma$ noise level) with 2.4 K km s$^{-1}$ intervals. The white lines show the observed area.
}
\label{fig:coha}
\end{figure}
\clearpage

\begin{figure}
\epsscale{0.4}
\plotone{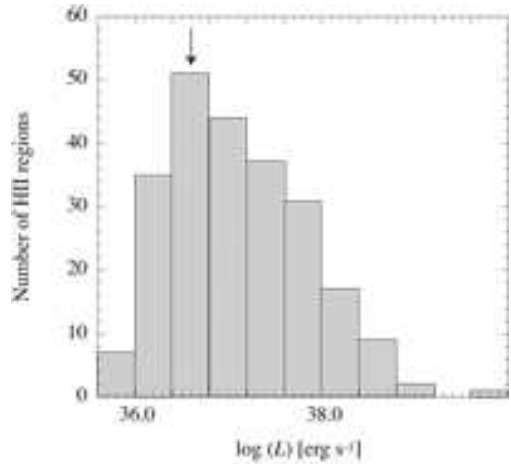}
\caption{
Distribution of the luminosities of the HII regions in the LMC 
from Kennicutt \& Hodge (1998).
The arrow shows the luminosity of the Orion nebula,
$L \sim 4 \times 10^{36}$ erg s$^{-1}$ (Gebel 1991).
}
\label{fig:HIIluminosity}
\end{figure}
\clearpage

\begin{figure}
\epsscale{0.4}
\plotone{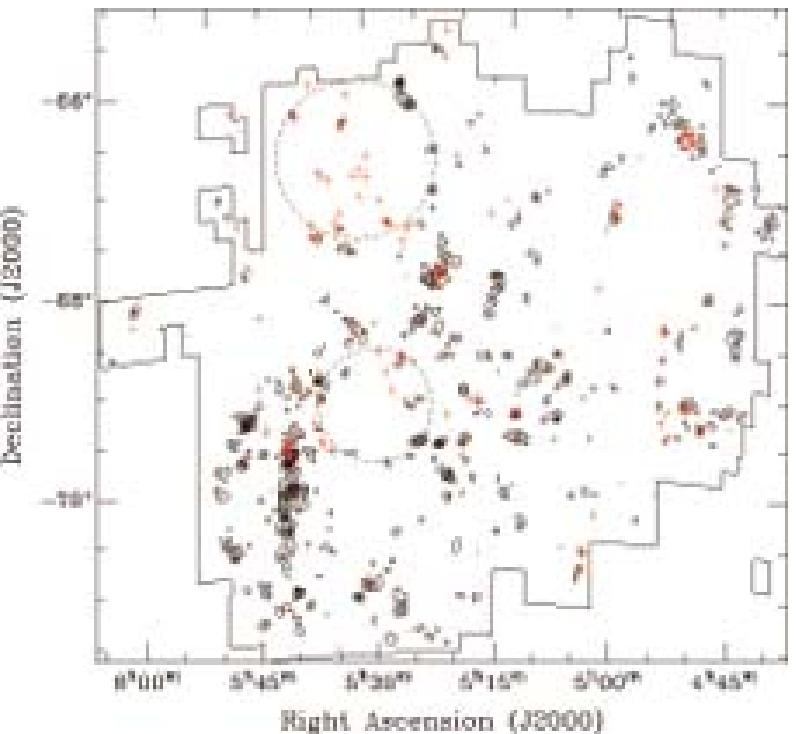}
\plotone{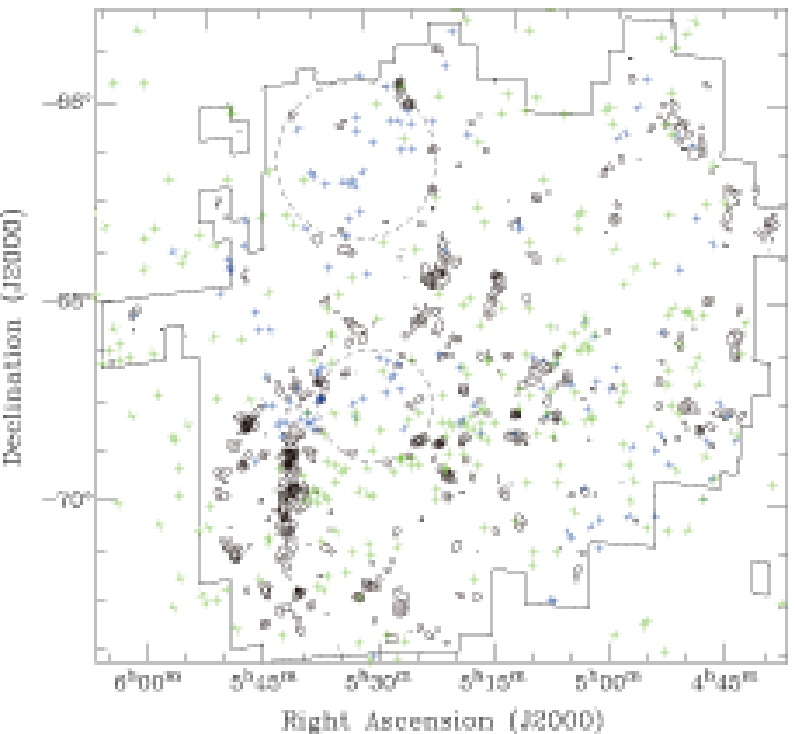}
\caption{Distribution of the clusters (Bica et al.\ 1996) and molecular clouds detected by NANTEN (Paper I). The contours show the velocity-integrated intensity of the CO (1--0) line obtained by NANTEN from 1.2 K km s$^{-1}$ (3 $\sigma$ noise level) with 2.4 K km s$^{-1}$ intervals. The thin lines show the observed area. The positions of the supergiant shells, LMC 3 and LMC 4, are shown by circles with dashed lines. The coodrdinates and the size are taken from Meaburn (1980). (a) Red crosses show the position of young clusters identified as SWB type 0 ($\tau \lesssim$ 10 Myr) by Bica et al.\ (1996). (b) Blue and green crosses show the positions of clusters identified as SWB type I (10 Myr $\lesssim \tau \lesssim$ 30 Myr) and SWB type II to VII (30 Myr $\lesssim \tau$), respectively.
}
\label{fig:coswb0}
\end{figure}
\clearpage

\begin{figure}
\epsscale{0.6}
\plotone{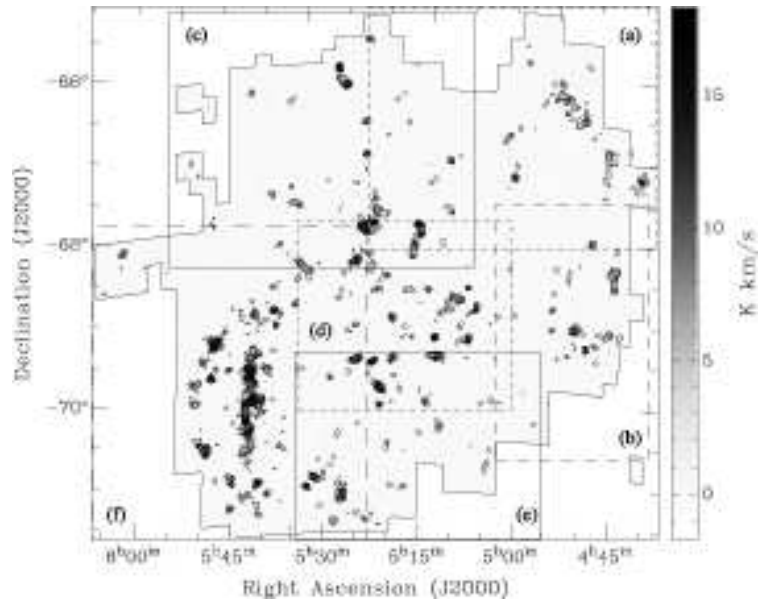}
\caption{
Velocity-integrated intensity map of the molecular clouds observed by NANTEN (Paper I). The contours are from 1.2 K km s$^{-1}$ (3 $\sigma$ noise level) with 1.2 K km s$^{-1}$ intervals. The boxes indicate the regions presented in Figures \ref{fig:e1_a} (a)--(f).
}
\label{fig:guide}
\end{figure}
\clearpage

\begin{figure*}
\epsscale{0.4}
\plotone{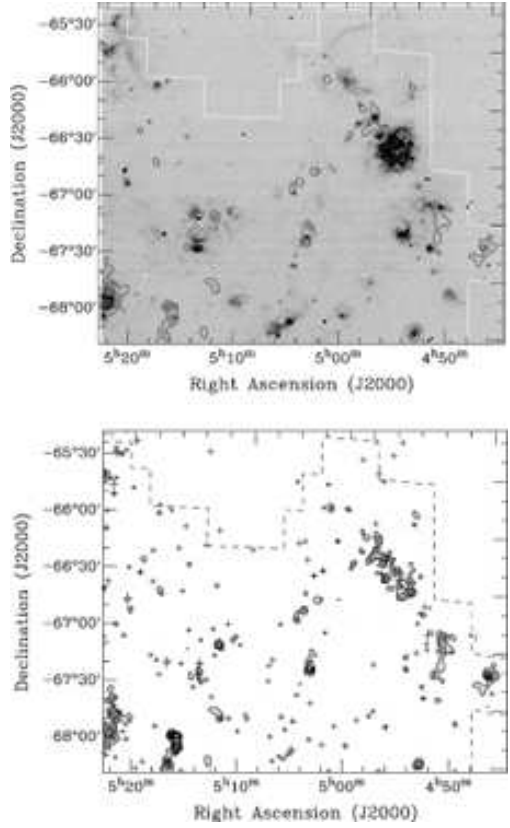}
\caption{
 ({\it upper panels} and {\it left panels} of Figures \ref{fig:e1_a} (a), (c)--(e), and \ref{fig:e1_a} (b), (f), respectively) Distribution of the molecular clouds (Paper I) superposed on the H $\alpha$ image (Kim et al. 1999); the contours are from 1.2 K km s$^{-1}$ (3 $\sigma$ noise level) with 2.4 K km s$^{-1}$ intervals. The white lines present the observed area.
 ({\it lower panels} and {\it right panels} of Figures \ref{fig:e1_a} (a), (c)--(e), and \ref{fig:e1_a} (b), (f), respectively) Distribution of the molecular clouds (Paper I, contours), HII regions by Henize (1956, yellow circles) and by Davies et al.\ (1976, red circles), youngest clusters identified as SWB type 0 (red crosses), SWB type I clusters (blue crosses) and SWB type II clusters (Bica et al.\ 1996, green crosses), respectively. The contours in red are from 1.2 K km s$^{-1}$ with 1.2 K km s$^{-1}$ intervals. The dashed lines present the observed area.
}
\label{fig:e1_a}
\end{figure*}
\clearpage
\begin{center}
\epsscale{0.5}
{\plotone{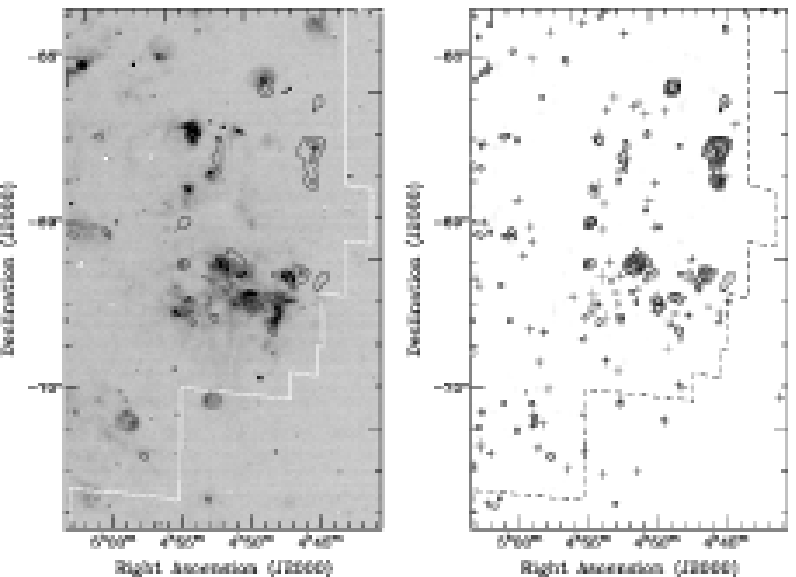}}\\[5mm]
\centerline{Fig. 5 (b). --- {\it continued}}
\epsscale{0.4}
{\plotone{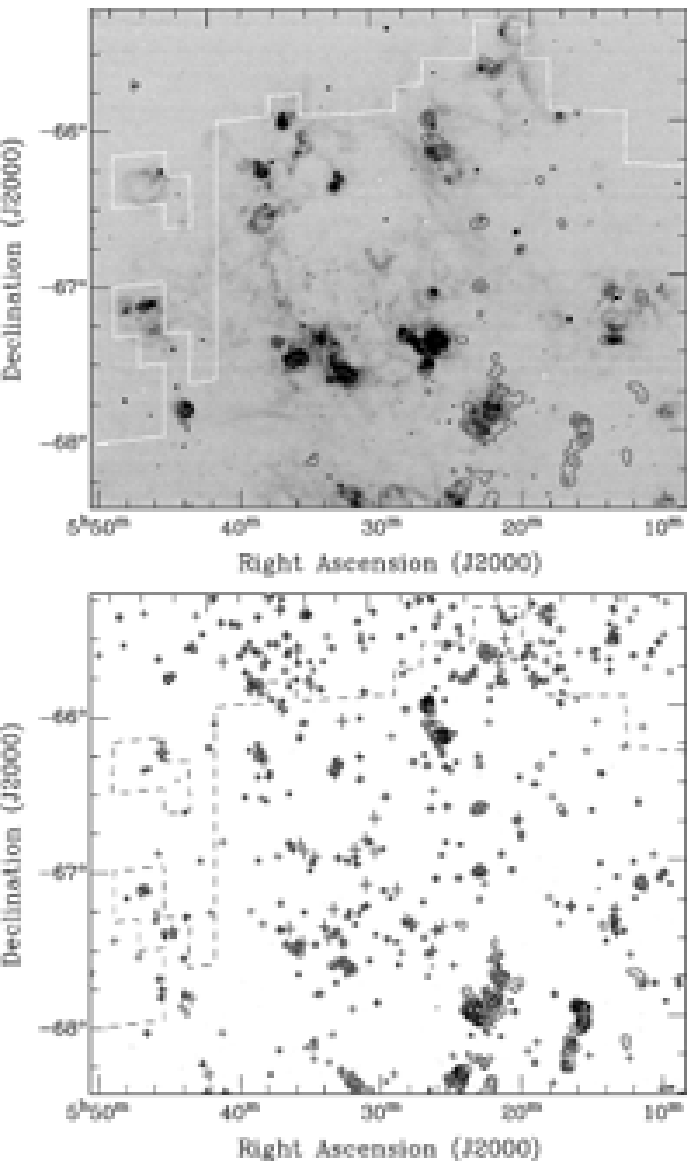}}\\[5mm]
\centerline{Fig. 5 (c). --- {\it continued}}
{\plotone{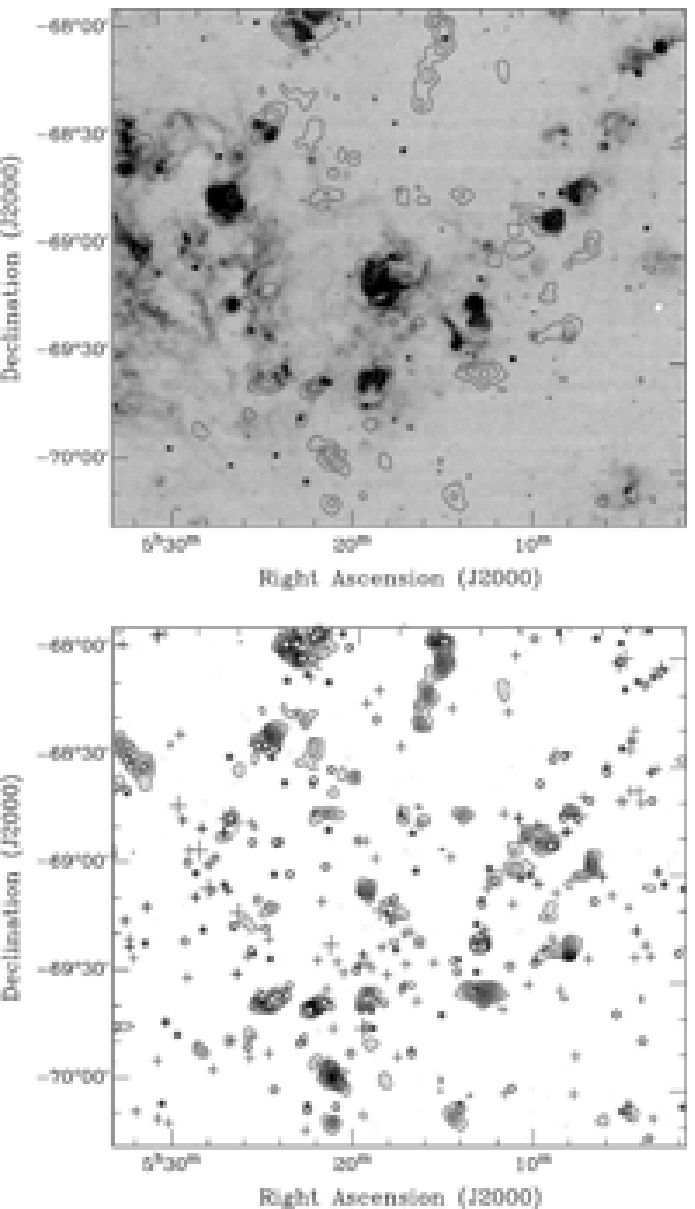}}\\[5mm]
\centerline{Fig. 5 (d). --- {\it continued}}
\clearpage
{\plotone{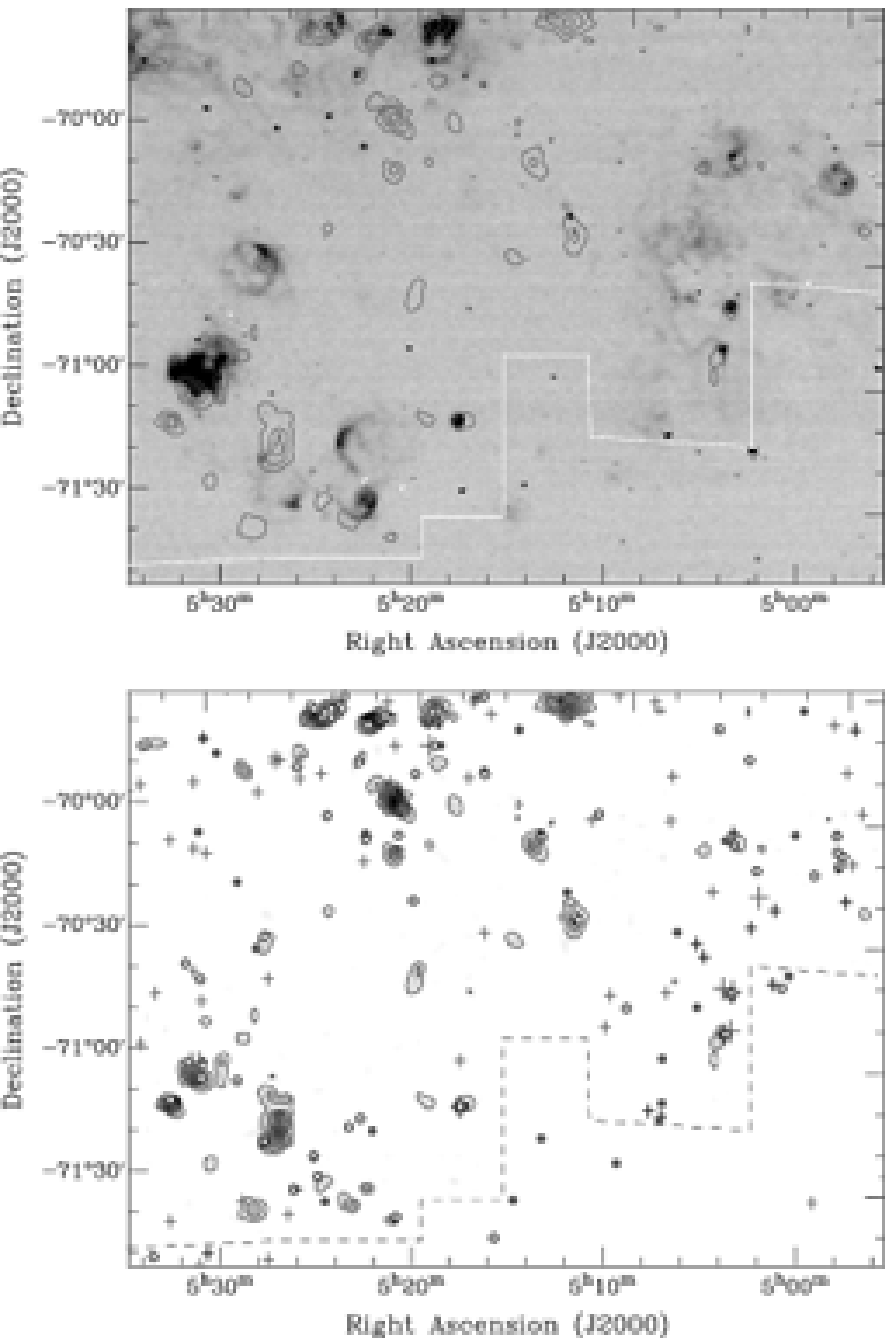}}\\[5mm]
\centerline{Fig. 5 (e). --- {\it continued}}
\epsscale{0.6}
{\plotone{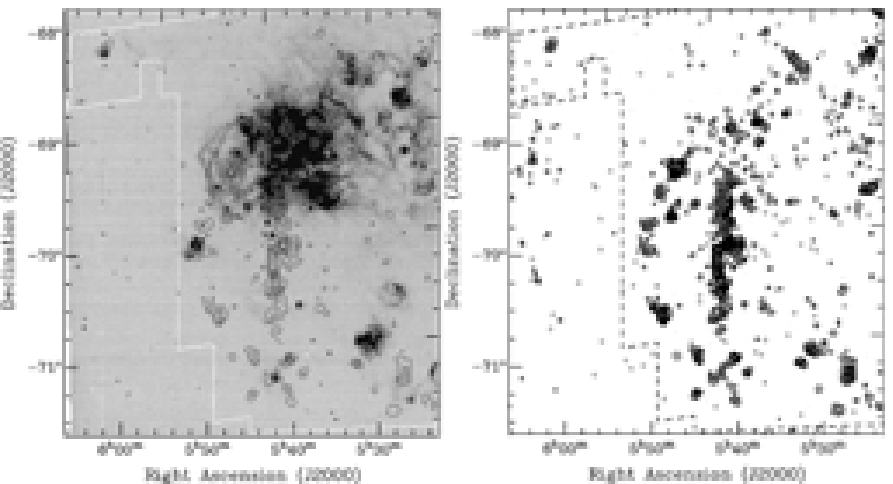}}\\[5mm]
\centerline{Fig. 5 (f). --- {\it continued}}
\end{center}
\clearpage

\begin{figure*}
\epsscale{0.5}
\plotone{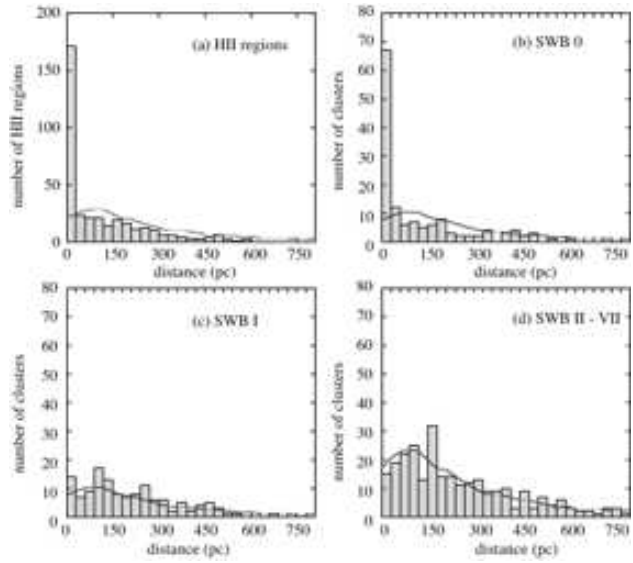}
\caption{
Frequency distribution of the projected distances of (a) the HII regions, (b) SWB 0 clusters ($\tau \lesssim$ 10 Myr), (b) SWB I clusters (10 Myr $\lesssim \tau \lesssim$ 30 Myr) and SWB type II to VII clusters (30 Myr $\lesssim \tau$, Bica et al. 1996) from the nearest molecular cloud (Paper I), respectively. Lines show the frequency distribution of the distance when the HII regions and clusters are distributed randomly.
}
\label{fig:dist}
\end{figure*}
\clearpage

\begin{figure*}
\epsscale{0.5}
\plotone{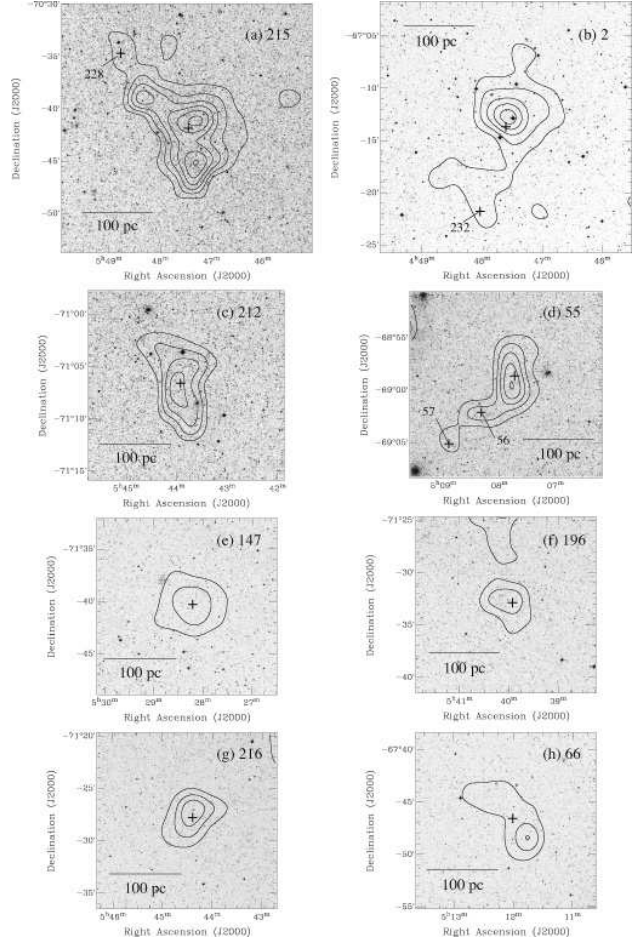}
\caption{
Examples of GMCs without massive star formation (GMC Type I). The eight most massive GMCs are shown in contours superposed on the DSS2 images. The contours are from 1.2 K km s$^{-1}$ with 1.2 K km s$^{-1}$ intervals. The crosses indicate the position of the GMCs as in Table 1 of Paper I. 
}
\label{fig:type1}
\end{figure*}

\clearpage
\begin{figure*}
\epsscale{0.5}
\plotone{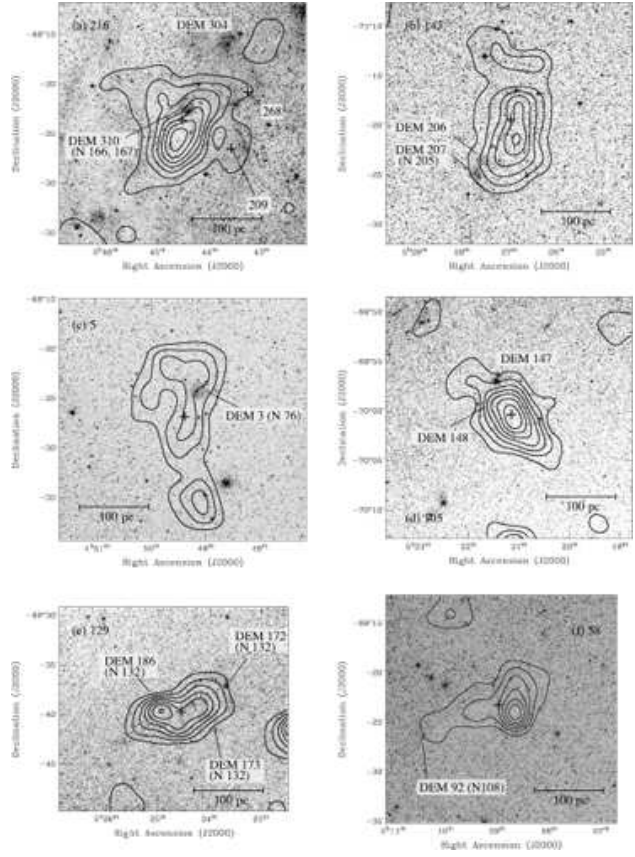}
\caption{
Examples of the molecular clouds associated with HII regions but not with young clusters (GMC Type II).
The six most massive GMCs are shown in contours superposed on the DSS2 images. The contours are from 1.2 K km s$^{-1}$ with 1.2 K km s$^{-1}$ intervals except for (a) GMC 216; the contours are from 1.2 K km s$^{-1}$ with 2.4 K km s$^{-1}$ intervals for (a). The crosses indicate the positions of the GMCs as in Table 1 of Paper I. 
}
\label{fig:type2}
\end{figure*}

\begin{figure*}
\epsscale{0.5}
\plotone{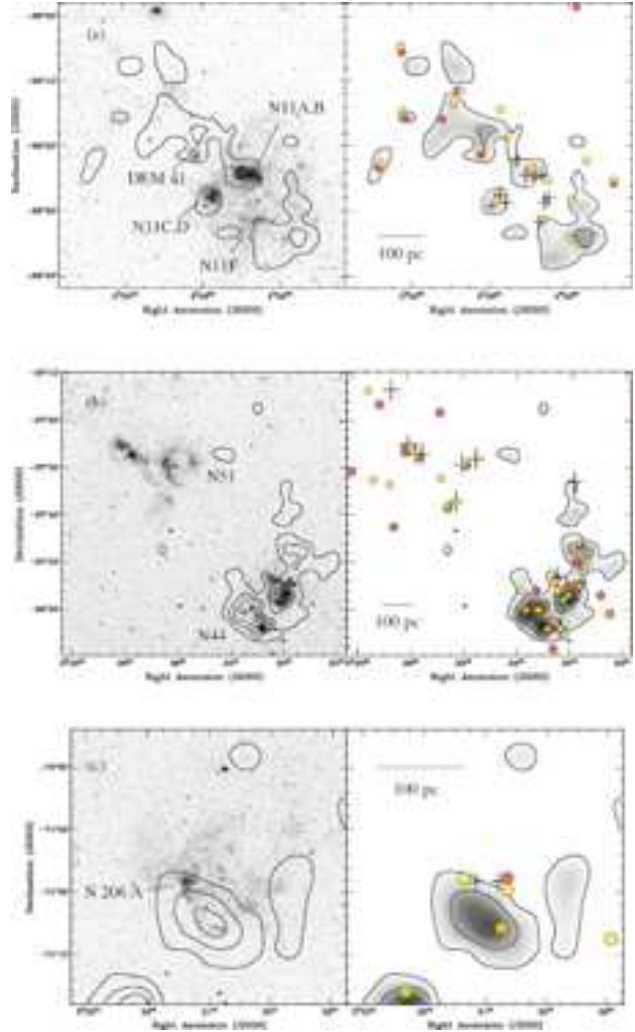}
\caption{
Examples of the molecular clouds associated with HII regions and young clusters
(GMC Type III) in (a) the N 11 region, (b) the N 44 and N 51 regions, (c) the N 206 region observed, and (d) the 30 Dor and N 159 regions.
({\it left panels}) Distribution of the molecular clouds by NANTEN (Paper I) is superposed on the H $\alpha$ image (Kim et al. 1999).
 ({\it right panels}) Distribution of the NANTEN molecular clouds (Paper I, contours),  HII regions (yellow circles by Henize et al.\ 1956 and red circles by Davies et al.\ 1976), and youngest clusters identified as SWB type 0 (Bica et al.\ 1996, crosses), respectively. The contours shown are from 1.2 K km s$^{-1}$ with 1.2 K km s$^{-1}$ intervals.
}
\label{fig:cluster1}
\end{figure*}
\clearpage
\epsscale{0.6}
{\plotone{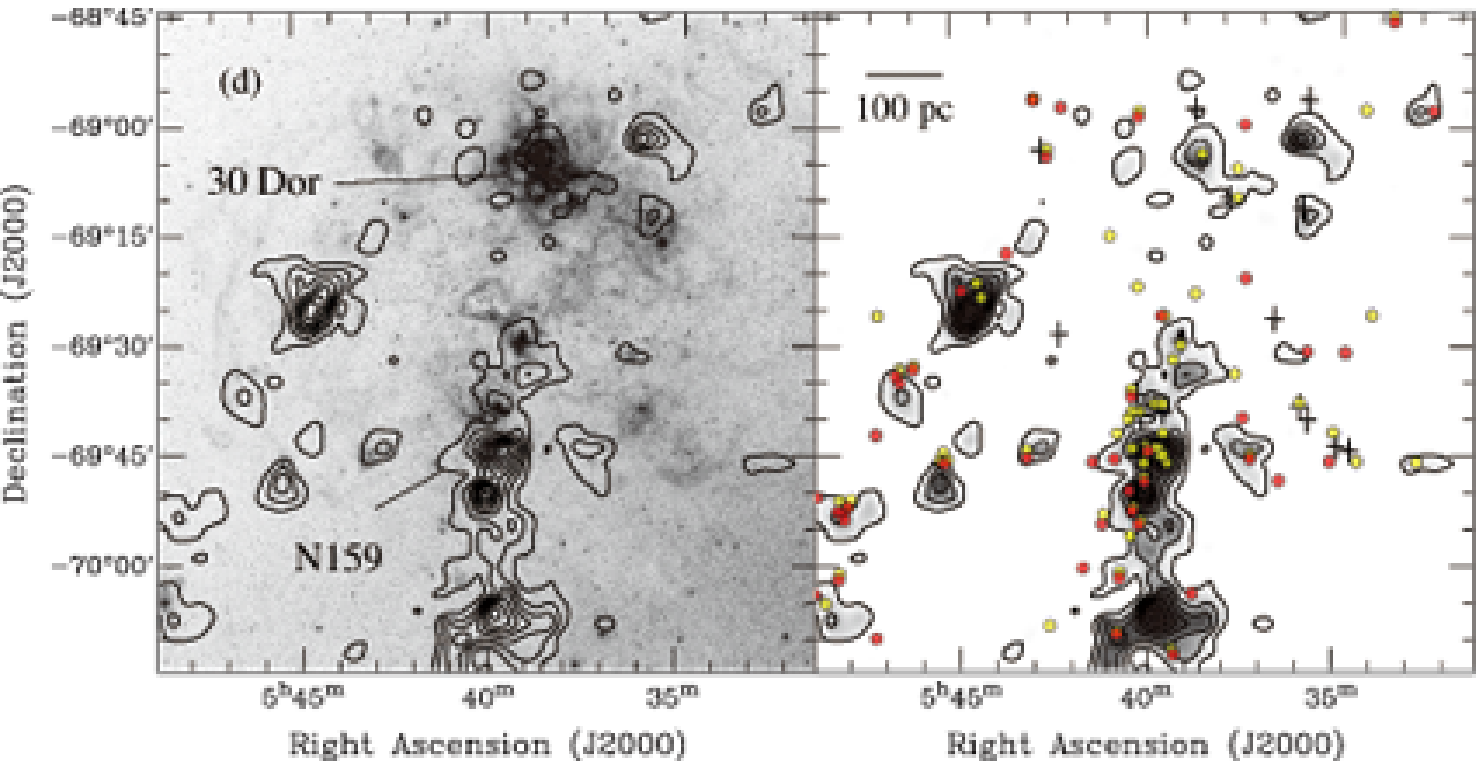}}\\[5mm]
\centerline{Fig. 9. --- {\it continued}}
\clearpage

\begin{figure*}
\epsscale{0.6}
\plotone{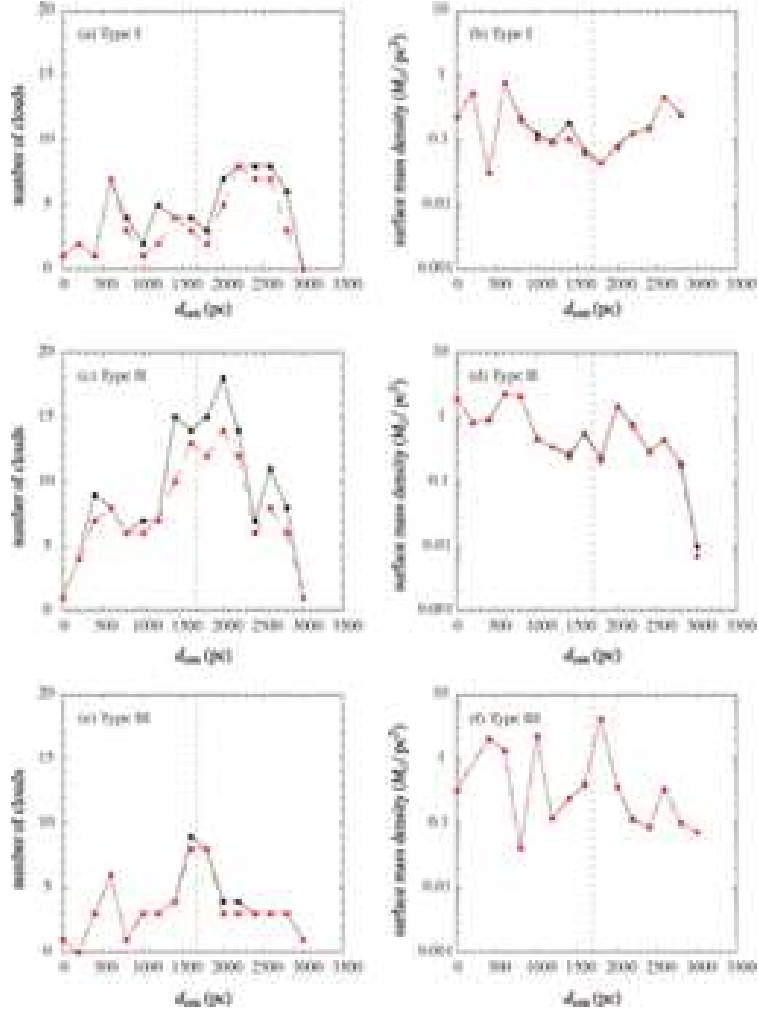}
\caption{
(a), (c), and (e) Frequency distribution of the cloud distances from the center 
$\alpha$(J2000) $= 5^h17.6^m$, $\delta$(J2000) $= -69\arcdeg 2\arcmin$
determined from the kinematics of the HI (Kim et al. 1998).
(b), (d), and (f) Distribution of the surface mass density along the distances from the center 
$\alpha$(J2000) $= 5^h17.6^m$, $\delta$(J2000) $= -69\arcdeg 2\arcmin$.
(a) and (b), (c) and (d), and (e) and (f) present those of the GMC Type I, Type II, and Type III, respectively.
Dashed lines [red in electric version] present those of the GMCs without small clouds.
The region within 1.7 kpc from the center (dotted lines) is completely covered (Paper I).
}
\label{fig:dcen}
\end{figure*}

\begin{figure*}
\epsscale{0.6}
\plotone{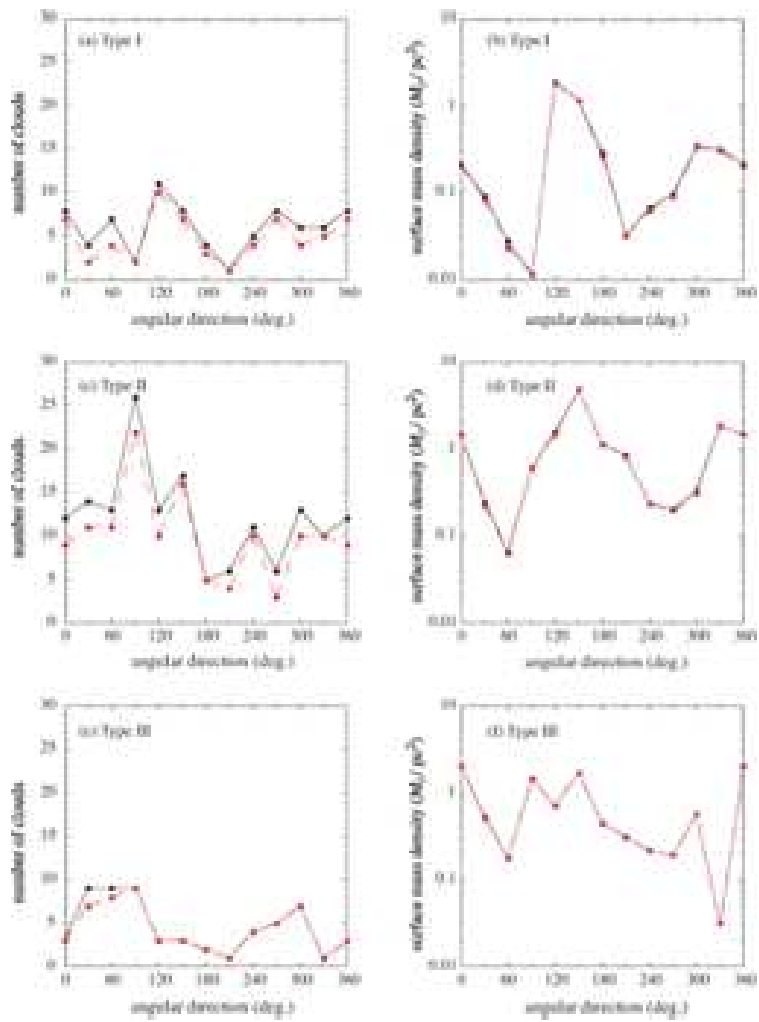}
\caption{
(a), (c), and (e) Frequency distribution of the position angle of the clouds with respect to the center 
$\alpha$(J2000) $= 5^h17.6^m$, $\delta$(J2000) $= -69\arcdeg 2\arcmin$
determined from the kinematics of the HI (Kim et al. 1998). 
The angle starts from the north to the east in counter clock-wise direction.
(b), (d), and (f) Distribution of the surface mass density along the position angle of the clouds with respect to the center $\alpha$(J2000) $= 5^h17.6^m$, $\delta$(J2000) $= -69\arcdeg 2\arcmin$.
(a) and (b), (c) and (d), and (e) and (f) present those of the GMC Type I, II, and III, respectively.
Dashed lines [red in electric version] present those of the GMCs without small clouds.
The region within 1.7 kpc from the center (dotted lines) is completely covered (Paper I).
}
\label{fig:angle}
\end{figure*}

\begin{figure*}
\epsscale{0.9}
\plotone{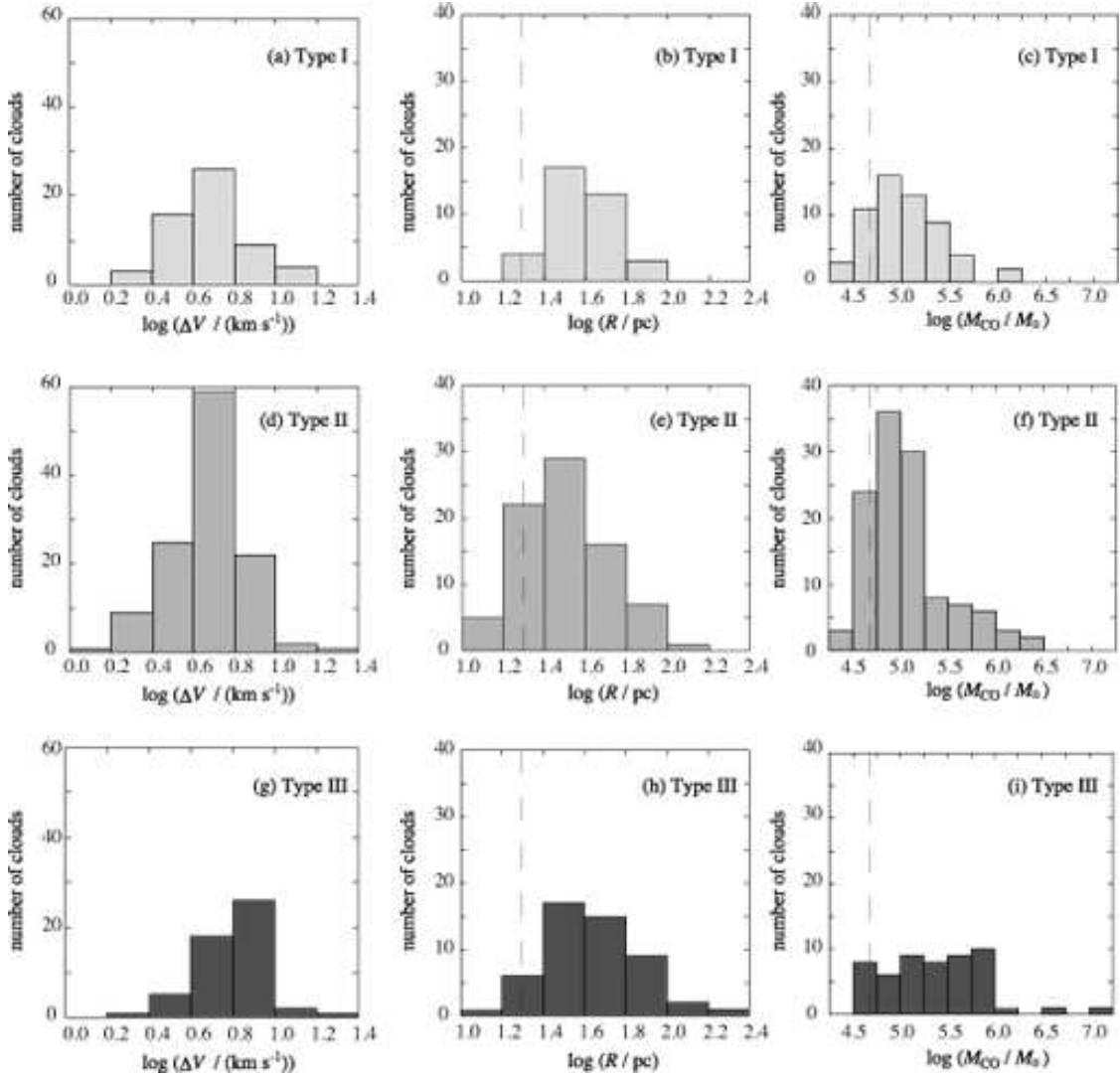}
\caption{
Histograms of $\Delta${\it V} ((a), (d), and (g)), {\it R} ((b), (e), and (h)), and $M_{\rm CO}$ ((c), (f), and (i)) of the GMCs. (a)--(c), (d)--(f), and (c)--(i) show the properties of the GMC Type I,
Type II, and Type III, respectively. Dashed lines indicate the completeness limit of the {\it R}, $R_{\rm completeness}=20$ pc, and $M_{\rm CO}$, $M_{\rm CO, completeness} = 5 \times 10^4 M_{\sun}$.  Note that {\it R} is derived for 164 GMCs out of 230 because the size cannot be derived for those with minor axis less than the NANTEN beam (see Paper I)
}
\label{fig:hist}
\end{figure*}

\begin{figure*}
\epsscale{0.5}
\plotone{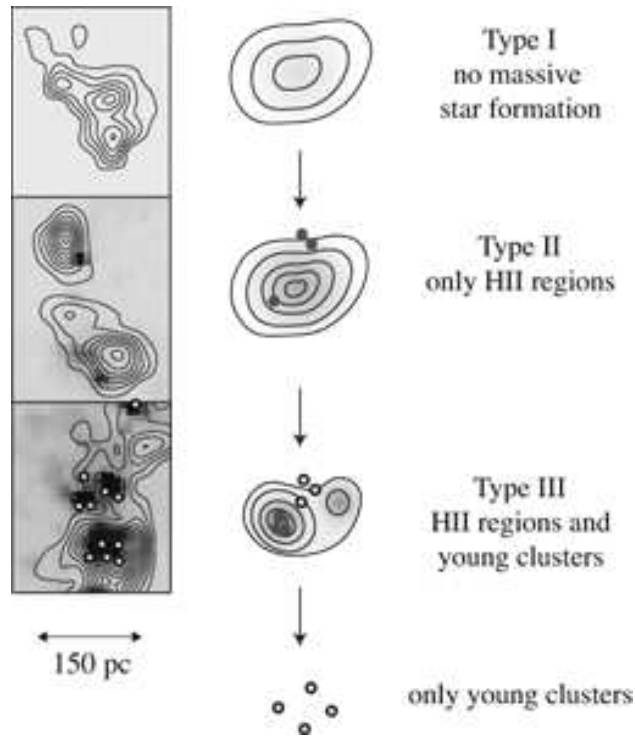}
\caption{
Evolutionary sequence of the molecular clouds.
The left panels are examples of GMC Type I (GMC 215, LMC N J0544-7127 in Table 1), II (GMC 135, LMC N J0525-6609), and III (the northern part of GMC 197, LMC N J0540-7008)  from the top panel,
respectively. Each panel presents H $\alpha$ images from Kim et al.\ (1999)
with GMCs identified by NANTEN (Paper I) in contours:
The contour levels are from 1.2 K km s$^{-1}$ with 1.2 K km s$^{-1}$
intervals. Open circles indicate the position of young clusters (Bica et al. 1996).
The middle panels are illustration for each evolutionary stage.
Open circles and filled circles in red represent young clusters and HII regions, respectively.
}
\label{fig:evolution}
\end{figure*}

\end{document}